\documentclass[useAMS,usenatbib,usegraphicx]{mn2e}

\usepackage{aas_macros}

\title[Accretion and magnetic fields around protostellar discs]{Accretion and magnetic field morphology around Class 0 stage protostellar discs}
  \author[D. Seifried et al.]
  {D.~Seifried,$^{1,2}$\thanks{seifried@ph1.uni-koeln.de} R.~Banerjee,$^{2}$ R.~E.~Pudritz,$^{3,4}$ R.~S.~Klessen$^{5,6,7}$ \\
  $^1$I. Physikalisches Institut, Universit\"at zu K\"oln, Z\"ulpicher Str. 77, 50937 K\"oln, Germany\\
  $^2$Hamburger Sternwarte, Universit\"at Hamburg, Gojenbergsweg 112, 21029 Hamburg, Germany\\
  $^3$Department of Physics $\&$ Astronomy, McMaster University, Hamilton ON L8S 4M1, Canada\\
  $^4$Origins Institute, McMaster University, ABB 241, Hamilton ON L8S 4M1, Canada\\
  $^5$Universit\"at Heidelberg, Zentrum f\"ur Astronomie, Institut f\"ur Theoretische Astrophysik, Albert-Ueberle-Str. 2, 69120 Heidelberg, Germany\\
  $^6$Department of Astronomy and Astrophysics, University of California, 1156 High Street, Santa Cruz, CA 95064, USA\\
  $^7$Kavli Institute for Particle Astrophysics and Cosmology, Stanford University, SLAC National Accelerator Laboratory, \\ Menlo Park, CA 94025, USA}

\date{Released 2014}

\pagerange{\pageref{firstpage}--\pageref{lastpage}} \pubyear{2013}

\begin{document}

\label{firstpage}

\maketitle

\begin{abstract}
We analyse simulations of turbulent, magnetised molecular cloud cores focussing on the formation of Class 0 stage protostellar discs and the physical conditions in their surroundings. We show that for a wide range of initial conditions Keplerian discs are formed in the Class 0 stage already. In particular, we show that even subsonic turbulent motions reduce the magnetic braking efficiency sufficiently in order to allow rotationally supported discs to form. We therefore suggest that already during the Class 0 stage the fraction of Keplerian discs is significantly higher than 50\%, consistent with recent observational trends but significantly higher than predictions based on simulations with misaligned magnetic fields, demonstrating the importance of turbulent motions for the formation of Keplerian discs. We show that the accretion of mass and angular momentum in the surroundings of protostellar discs occurs in a highly anisotropic manner, by means of a few narrow accretion channels. The magnetic field structure in the vicinity of the discs is highly disordered, revealing field reversals up to distances of 1000 AU. These findings demonstrate that as soon as even mild turbulent motions are included, the classical disc formation scenario of a coherently rotating environment and a well-ordered magnetic field breaks down. Hence, it is highly questionable to assess the magnetic braking efficiency based on non-turbulent collapse simulation. We strongly suggest that, in addition to the global magnetic field properties, the small-scale accretion flow and detailed magnetic field structure have to be considered in order to assess the likelihood of Keplerian discs to be present.
\end{abstract}


\section{Introduction}

The impact of magnetic fields on the formation of protostellar discs in the Class 0 phase has received great attention in the last decade \citep[see e.g.][for a recent review]{Li14}. In previous works a highly idealised numerical setup usually consisting of a rotating core and a magnetic field parallel to the rotation axis was often used \citep[e.g.][]{Allen03,Matsumoto04,Machida05,Banerjee06,Banerjee07,Price07,Hennebelle08,Hennebelle09,Duffin09,Commercon10,Burzle11,Seifried11}. A common result of these simulations is that for a magnetic field strength comparable to observations, i.e. a mass-to-flux ratio smaller than 5 -- 10 \citep[e.g.][]{Falgarone08,Girart09,Beuther10}, no rotationally supported discs were formed. This rather artificial result owing to the idealised initial conditions (see below) was called the ''magnetic braking catastrophe'' owing to the fact that magnetic braking is responsible for the removal of the angular momentum which would be necessary to form the discs.

These numerical findings, however, seem to contradict the results of a growing number of high-resolution observations which strongly indicate the existence of Keplerian discs around Class 0 protostellar objects \citep{Tobin12,Hara13,Murillo13,Sanchez13,Codella14}. By observations of line transition emission the authors could show that the velocity profile around the observed Class 0 stage objects agrees with that of a Keplerian disc up to radii of $\sim$ 100 AU. Also previous observations based on SED modelling suggest the presence of well-defined, rotationally supported discs \citep{Jorgensen09,Enoch11}. In summary, recent observations strongly indicate the presence of Keplerian discs in contrast to the simulations mentioned in the beginning.

One approach to overcome the apparent conflict between observational and simulation results was the inclusion of non-ideal MHD effects. Ambipolar diffusion~\citep[e.g.][]{Mellon09,Duffin09}, however, does not result in the formation of Keplerian discs in the earliest phase of protostellar evolution. Furthermore, including Ohmic dissipation seems to allow only for very small ($\sim$ 10 solar radii) rotationally supported structures \citep[e.g.][]{Krasnopolsky10,Dapp12}, and even the combined effects of ambipolar diffusion and Ohmic dissipation was found not to reduce the magnetic braking efficiency significantly~\citep{Li11}. However, in more recent work \citet{Machida14} find Keplerian discs to appear in the presence of Ohmic dissipation, although the results appear to strongly depend on the numerical settings. Hence, it seems that further work in this field is required to clarify the impact of non-ideal MHD effects.

On the other hand, \citet{Hennebelle09}, \citet{Ciardi10} and \citet{Joos12} performed simulations deviating from the highly idealised setup of a uniformly rotating core and a magnetic field parallel to the rotation axis were performed. The authors could show that for an overall magnetic field inclined to the core's rotation axis the formation of rotationally supported discs is possible. However, based on a realistic distribution of magnetic field strengths and misalignment angles, \citet{Krumholz13} pointed out that this would lead to a fraction of Keplerian discs in the Class 0 stage of only 10 to at most 50\%, and thus noticeably below the observed fraction of discs around Class I/II objects \citep[e.g.][]{Williams11}.

In a different approach, the inclusion of turbulent motions in the initial conditions and their effect on the formation of rotationally supported discs was studied~\citep{Seifried12,Seifried13,Santos12,Santos13,Joos13,Myers13,Li14b}. The authors showed that turbulence enables the formation of Keplerian discs for a wide range of conditions even in the presence of strong magnetic fields. \citet{Santos12,Santos13} attribute the formation of rotationally supported discs to turbulent reconnection occurring in the vicinity of the disc and an associated loss of magnetic flux, which in turn reduces the magnetic braking efficiency. In contrast, in~\citet{Seifried12,Seifried13} we suggest that the build-up of Keplerian discs is due to the turbulent motions and the disordered magnetic field structure in the surroundings of the discs rather than due to magnetic flux loss. The turbulent motions distort the magnetic field and hamper the build-up of a toroidal magnetic field component responsible for angular momentum extraction. Simultaneously they provide a sufficient amount of angular momentum necessary for the formation of the disc.

In this work we analyse in detail the physical conditions in the vicinity of protostellar discs which are responsible for the Keplerian rotation structure. We are in particular interested in how accretion towards and onto the discs occurs and how the magnetic field morphology is affected by the turbulent motions. Moreover, we investigate whether in massive molecular cloud cores already subsonic turbulent motions are sufficient to allow the -- as we call it -- turbulence-induced disc formation to set in. For low-mass cores we have already demonstrated that subsonic turbulence is sufficient to allow Keplerian to form \citet{Seifried13}. Furthermore, we study if or to what extent the conditions in the surroundings of the protostellar discs are different in the presence or absence of a global core rotation. A main outcome of this work is that under realistic conditions, i.e. as soon as even weak turbulent motions are included, the accretion in the vicinity of protostellar discs is highly anisotropic and that the magnetic field structure is highly disordered.

The paper is organised as follows. In Section~\ref{sec:IC} we briefly describe the initial conditions of the simulations and summarise the main results of our previous work \citep{Seifried12,Seifried13}. The results concerning the accretion and magnetic field structure are presented in Section~\ref{sec:longterm} focussing in detail on two fiducial simulations. In Section~\ref{sec:ICdependence} we extend this analysis to a wider range of initial conditions and investigate under which conditions Keplerian discs are formed. In Section~\ref{sec:discussion} the results are discussed in a broader context and are compared to related numerical and observational work before we summarise our main findings in Section~\ref{sec:conclusions}.

\section{Initial conditions and basic results}
\label{sec:IC}

The simulations presented here are performed with the astrophysical code FLASH \citep{Fryxell00} in version 2. We make use of the sink particle routine \citep{Federrath10} to model the formation of protostars. The results of the simulations were first presented in \citet{Seifried12,Seifried13}. We here only briefly describe the basic setup of the simulations, for more details we refer the reader to the aforementioned papers. In the simulations we consider the collapse of turbulent, magnetised molecular cloud cores and the subsequent formation and evolution of protostellar discs. The cores have masses of 2.6 M$_{\sun}$ and 100 M$_{\sun}$ and a radius of 0.0485 pc and 0.125 pc, respectively (see Table~\ref{tab:models} for an overview over the different models and their nomenclature). The density profile of the low-mass core is that of Bonnor-Ebert sphere, scaled up to contain the 2.6 M$_{\sun}$, whereas the profile of the 100 M$_{\sun}$ declines outwards with the radius as $\rho \propto r^{-1.5}$. The temperature of the low-mass cores is set to 15 K, and that of the high-mass cores to a somewhat higher value of 20 K \citep[see e.g.][]{Ragan12,Launhardt13}. The cores are embedded in an ambient low-density medium in pressure equilibrium with the edge of the cores.
\begin{table}
 \caption{Initial conditions of the performed simulations showing the mass and radius of the molecular cloud core, whether uniform rotation is present or not, the rms Mach number, and the simulated evolutionary times of the protostellar discs.}
 \label{tab:models}
 \begin{tabular}{@{}lccccc}
  \hline
  Run & $m_\rmn{core}$ & r$_\rmn{core}$ & rotation & $M_\rmn{rms}$ & t$_\rmn{sim}$ \\
      &  [M$_{\sun}$]  & [pc]           &          &               & [kyr] \\
  \hline
  M2-NoRot & 2.6 & 0.0485 &  No & 0.74 & 30 \\
  M100-NoRot & 100 & 0.125 &  No & 2.5 & 25 \\
  M100-NoRot-Mrms0.5 & 100 & 0.125 &  No & 0.5 & 10 \\
  M100-NoRot-Mrms1 & 100 & 0.125 &  No & 1.0 & 10 \\
  M100-NoRot-Mrms2.5 & 100 & 0.125 &  No & 2.5 & 12.5 \\
  M2-Rot & 2.6 & 0.0485 &  Yes & 0.74 & 15 \\
  M100-Rot & 100 & 0.125 &  Yes & 2.5 & 20 \\
  \hline
 \end{tabular}
\end{table}

The cores are threaded by a magnetic field parallel to the $z$-axis with the strength chosen such that the (normalised) mass-to-flux ratio $\mu$ of the cores is 2.6. We emphasise that throughout the paper the mass-to-flux ratio is given in units of the critical mass-to-flux ratio $\mu_\rmn{crit} = 0.13/\sqrt{G}$ \citep{Mouschovias76}, where $G$ is the gravitational constant.

In all cores a Kolmogorov-type turbulent velocity field is present with the 3D-turbulent rms Mach number ($M_\rmn{rms}$) ranging from subsonic to supersonic values. The runs M100-NoRot and M2-NoRot are identical to the simulations already considered in the previous papers but are followed over an even longer timescale of 25 kyr (run M100-NoRot) and 30 kyr (run M2-NoRot) after the formation of the first protostar\footnote{We note that in the simulations we have neglected radiative feedback of the protostars.}. We point out that in neither of the two cases there is a global rotation present in the beginning. For comparative purposes we also consider two runs with identical initial conditions in which global rotation is superimposed on the turbulence field (the runs M100-Rot and M2-Rot). In these two runs, the rotation frequency is chosen such that the rotation energy equals the turbulent kinetic energy, i.e. $\Omega = 3.16 \cdot 10^{-13}$ s$^{-1}$ and $2.2 \cdot 10^{-13}$ s$^{-1}$ for run M100-Rot and M2-Rot, respectively. For the case of a 100 M$_{\sun}$ core we have performed three further simulations with varying turbulence strength in order to test the dependence of the turbulence-induced disc formation mechanism proposed in \citet{Seifried12,Seifried13} on the actual strength of the turbulence (runs M100-NoRot-Mrms0.5, M100-NoRot-Mrms1, and M100-NoRot-Mrms2.5). For these three runs we have used different initial turbulence seeds with $M_\rmn{rms}$ of 0.5, 1, and 2.5.

The cooling routine applied in our runs takes into account dust cooling, molecular line cooling and the effects of optically thick gas \citet{Banerjee06}. We introduce sink particles above a density threshold of $\rho_\rmn{crit} = 1.14 \cdot 10^{-10}$ g cm$^{-3}$ using an accretion radius of 3.14 AU \citep[see][for details]{Federrath10}. Here we only point out that the angular momentum carried by the gas will be transferred to the spin of the sink particle. The magnetic field is left unchanged during the accretion process in order to avoid the violation of the divergence-free condition.
The maximum spatial resolution in our simulations is set to 1.2 AU. We apply a refinement criterion which guarantees that the Jeans length is resolved everywhere with at least eight grid cells.

Next, we briefly recapitulate the main outcomes of the runs M100-NoRot, M2-NoRot, M100-Rot, and M2-Rot already discussed in \citet{Seifried12,Seifried13}. We find that Keplerian discs are formed in all four simulations independently of the core mass, the strength of turbulence, or the presence/absence of global rotation. The discs form within a few kyr after the formation of the protostar, are 50 -- 150 AU in size, and have masses of 0.05 up to a few 0.1 M$_{\sun}$. We find that in the initial phase of disc formation the mass-to-flux ratio stays well below a value of 10. From non-turbulent simulations this value was found to be the critical value for Keplerian disc formation, below which no rotationally supported discs were found \citep{Allen03,Matsumoto04,Machida05,Banerjee06,Banerjee07,Price07,Hennebelle08,Hennebelle09,Duffin09,Commercon10,Seifried11}. Hence, we conclude that flux-loss alone cannot explain the formation of Keplerian discs in our case. In \citet{Seifried12,Seifried13} we argued that the formation of rotationally supported discs at such early phases is rather due to the disordered magnetic field structure and due to shear flows created by the turbulent motions in the surroundings of the discs, which carry significant amounts of angular momentum. We concluded that these two effects lower the classical magnetic braking efficiency compared to a highly idealised rotating system with a magnetic field of comparable strength but perfectly aligned with the rotation axis.

In the following we analyse the complex and disordered morphology of the magnetic field, the occurrence of shear flows and the associated anisotropic accretion in a more quantitative way than done in our previous work.

\section{Long-term evolution of protostellar discs}
\label{sec:longterm}

We first consider the runs M2-NoRot and M100-NoRot already discussed in \citet{Seifried12,Seifried13}. We let the protostellar system evolve such that the simulations cover a time of 30 kyr (run M2-NoRot) and 25 kyr (run M100-NoRot) after the formation of the first protostar. The total protostellar masses are 0.62 and 15.65 M$_{\sun}$ for run M2-NoRot and M100-NoRot, respectively, i.e. 24\% and 16\% of the initial core mass. We note that for run M2-NoRot only one protostar has formed whereas for run M100-NoRot 8 protostars have formed with the most massive one (which is also the first to have formed) having a mass of 9.26 M$_{\sun}$.

\begin{figure*}
 \includegraphics[height=0.45\linewidth]{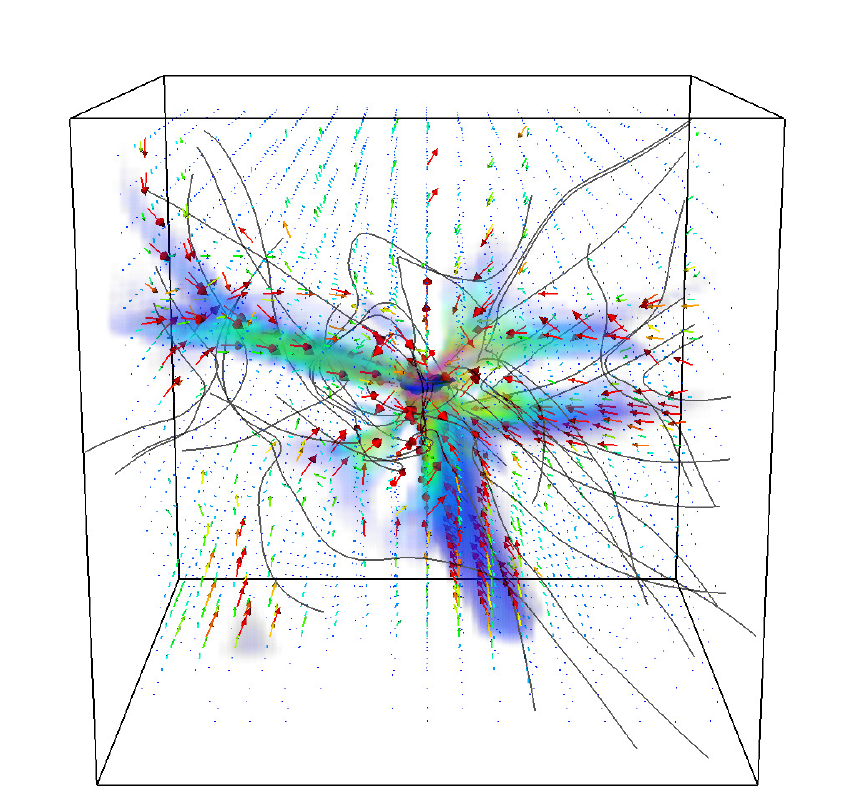}
 \includegraphics[height=0.45\linewidth]{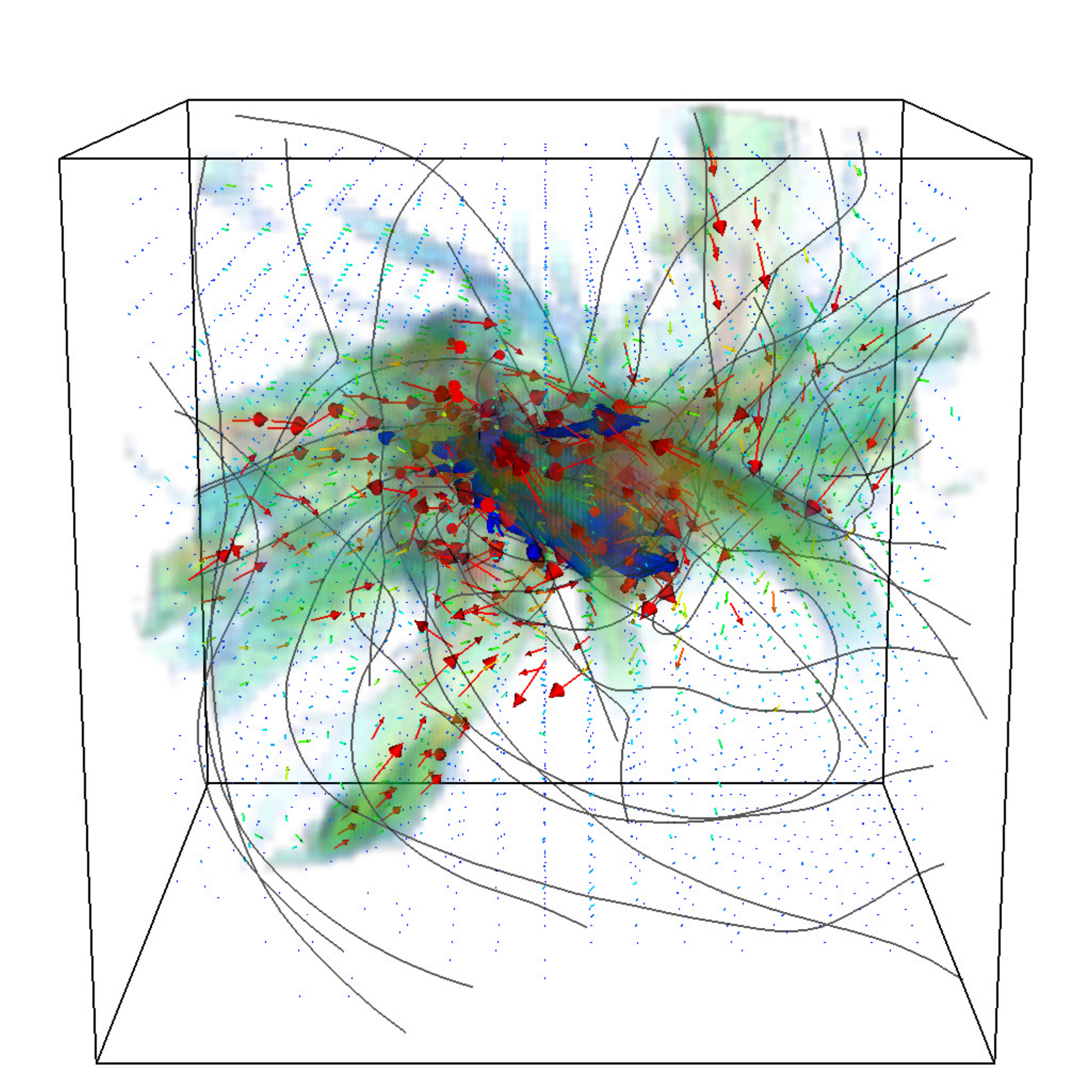}
\caption{Structure of the accretion flow $\dot{m}$ (vectors and coloured volume rendering, see Section~\ref{sec:accrflow} for details of the calculation) and the magnetic field (black lines) around the protostellar disc (blue contour) at the end of run M2-NoRot (left) and run M100-NoRot (right). The magnetic field structure is highly disordered and accretion occurs preferentially along a few narrow accretion channels. The box is 1300 AU in size.}
\label{fig:overview}
\end{figure*}

In both runs a first protostar forms in an overdense condensation. We emphasise that at the very moment the protostar forms no rotationally supported structure is recognisable. Subsequently, due to ongoing accretion of angular momentum the Keplerian discs builds up within a few kyr. The disc remains Keplerian over the entire evolution, unless a violent accretion event disturbs the disc. In this case, however, a rotationally supported structure quickly reestablishes. At the end of both runs the protostellar discs show a clear Keplerian rotation structure and only mild infall motions \citep[see Fig. 5 in][]{Seifried13}. In Fig.~\ref{fig:overview} we display the situation at the end of both runs showing the magnetic field structure and the accretion flow in the vicinity of the protostellar disc. One can already recognise that the magnetic field structure is highly complex. The field lines are strongly bent or even completely reversed in their direction. Moreover, also the accretion flow morphology, displayed by the coloured regions and vectors, reveals a highly anisotropic structure. In the following we analyse both the magnetic field structure and the accretion flow more quantitatively.

\subsection{Low-mass core simulation}

We first consider the results of the collapse of a low-mass (2.6 M$_{\sun}$), non-rotating core (run M2-NoRot). We describe in detail the calculations made to quantify the degree of disorder of the magnetic field and the anisotropy of the accretion flow. We start by considering the accretion flow before we analyse the magnetic field structure.

\subsubsection{Accretion flow}
\label{sec:accrflow}

Here we analyse the accretion flow towards/onto the protostellar disc formed in run M2-NoRot. For this purpose in a first step we determine the disc's centre of mass and centre-of-mass velocity. For the calculation of the disc properties we only take into account gas with a density larger than $5 \cdot 10^{-13}$ g cm$^{-3}$. Next, we subtract the disc's centre-of-mass velocity from the gas velocity in order to obtain the overall gas motions relative to the disc centre\footnote{We note that over the entire time the disc's centre of mass and the sink position(s) agree within a few AU. Hence, we safely can neglect the particles centre-of-mass velocity and only take the disc's centre of mass velocity when calculating the relative gas motions.}. Finally, we take the radial component of the (relative) gas motions, which we multiply with the gas density in order to properly describe the \textit{mass} accretion flow towards (or away from) the disc. Hence, the accretion flow is defined as
\begin{equation}
 \dot{m} = \rho \cdot v_\rmn{radial}
\end{equation}
which is technically a \textit{mass accretion rate per area}. This accretion flow is visualised in Fig.~\ref{fig:overview} revealing a quite complex structure. For the accretion of the \textit{material angular momentum} $\dot{l}_{\rmn{gas}}$ we multiply the mass accretion flow with the specific angular momentum of the gas with respect to the disc's angular momentum $\bmath{L}_{\rmn{disc}}$, i.e. 
\begin{equation}
 \dot{l}_ {\rmn{gas}} = \dot{m} \cdot (\bmath{r} \times \bmath{v}) \cdot \bmath{L}_{\rmn{disc}}/|\bmath{L}_{\rmn{disc}}| \, .
\end{equation}

We study the accretion flow through spheres with radii of 50, 250, 500, and 1000 AU around the centre of the disc. We emphasise that all data shown in the following are averaged over a spherical shell with a thickness of 10 AU.
\begin{figure*}
\centering
 \includegraphics[width=0.24\linewidth]{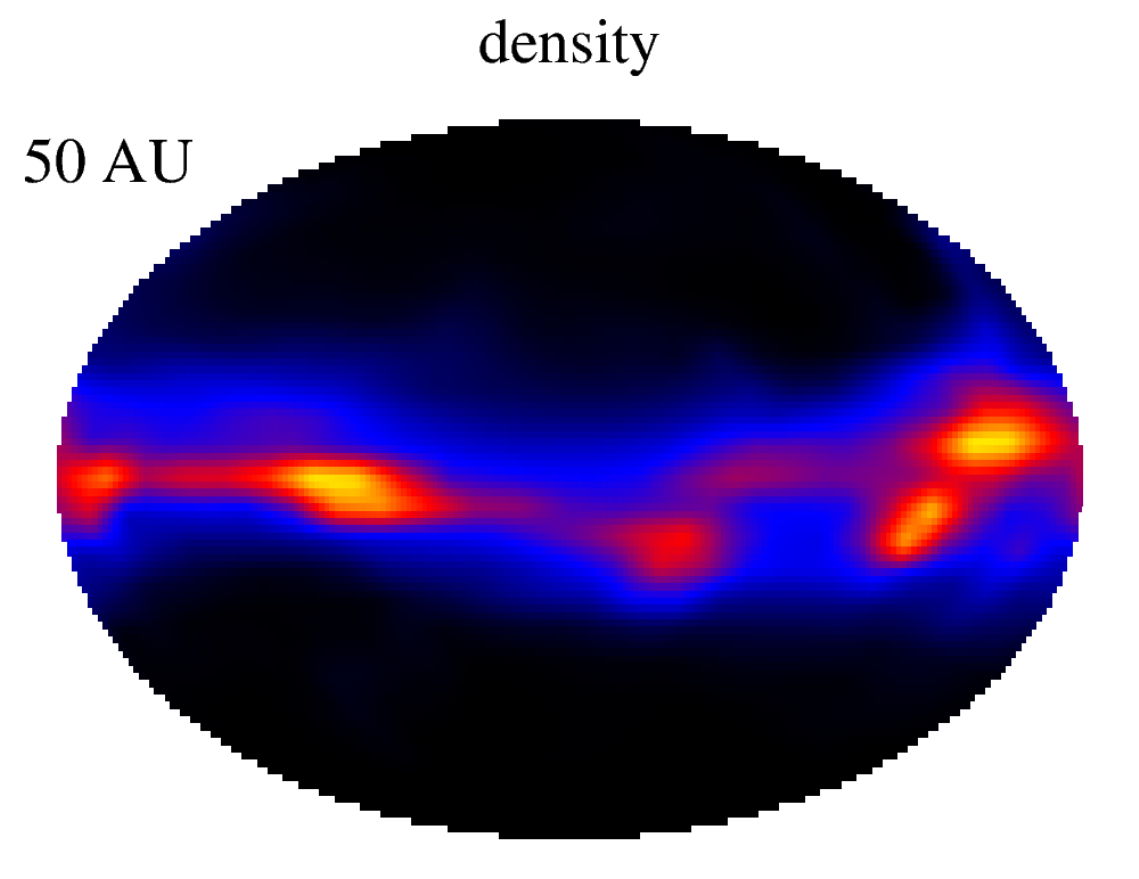}
 \includegraphics[width=0.24\linewidth]{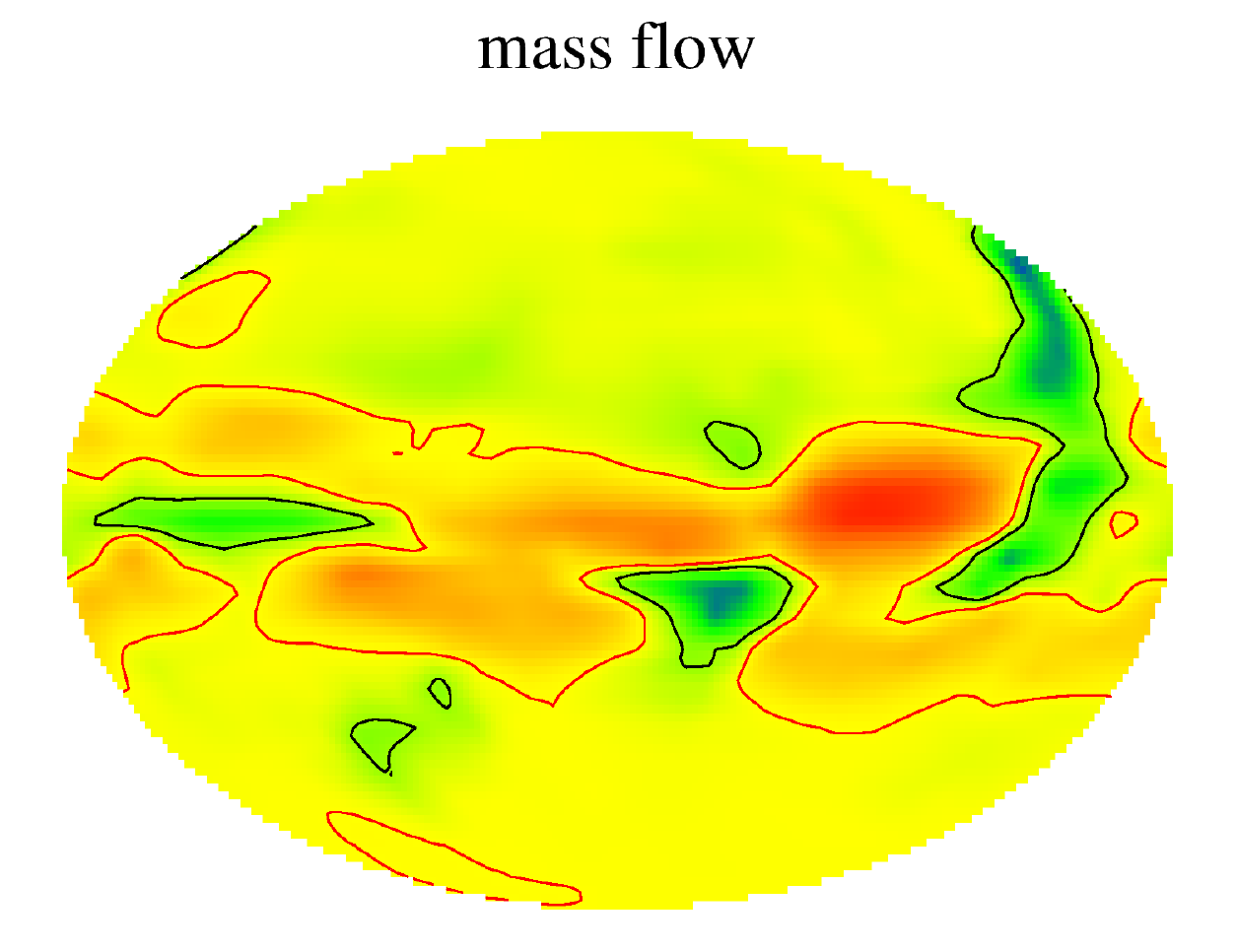}
 \includegraphics[width=0.24\linewidth]{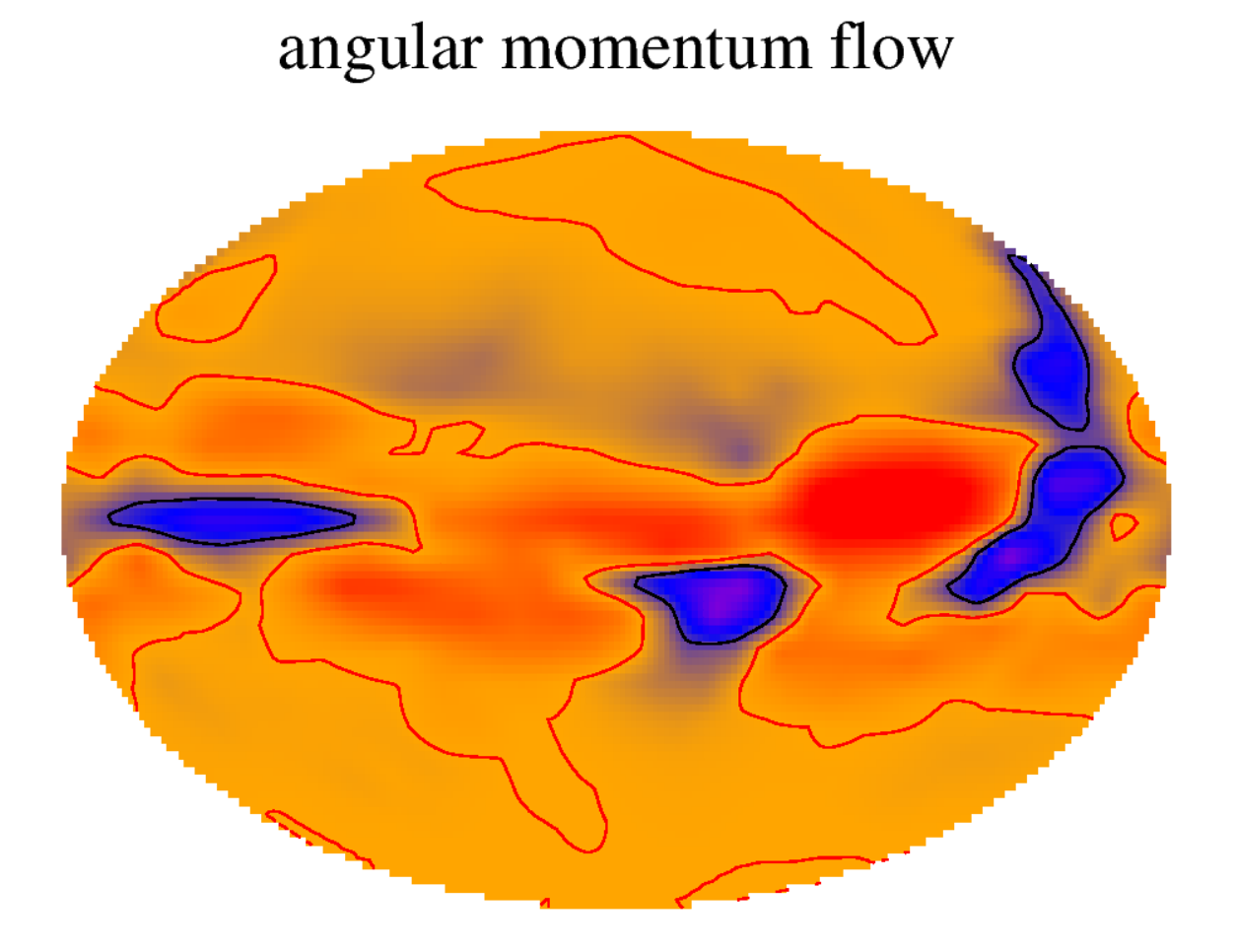} 
 \includegraphics[width=0.24\linewidth]{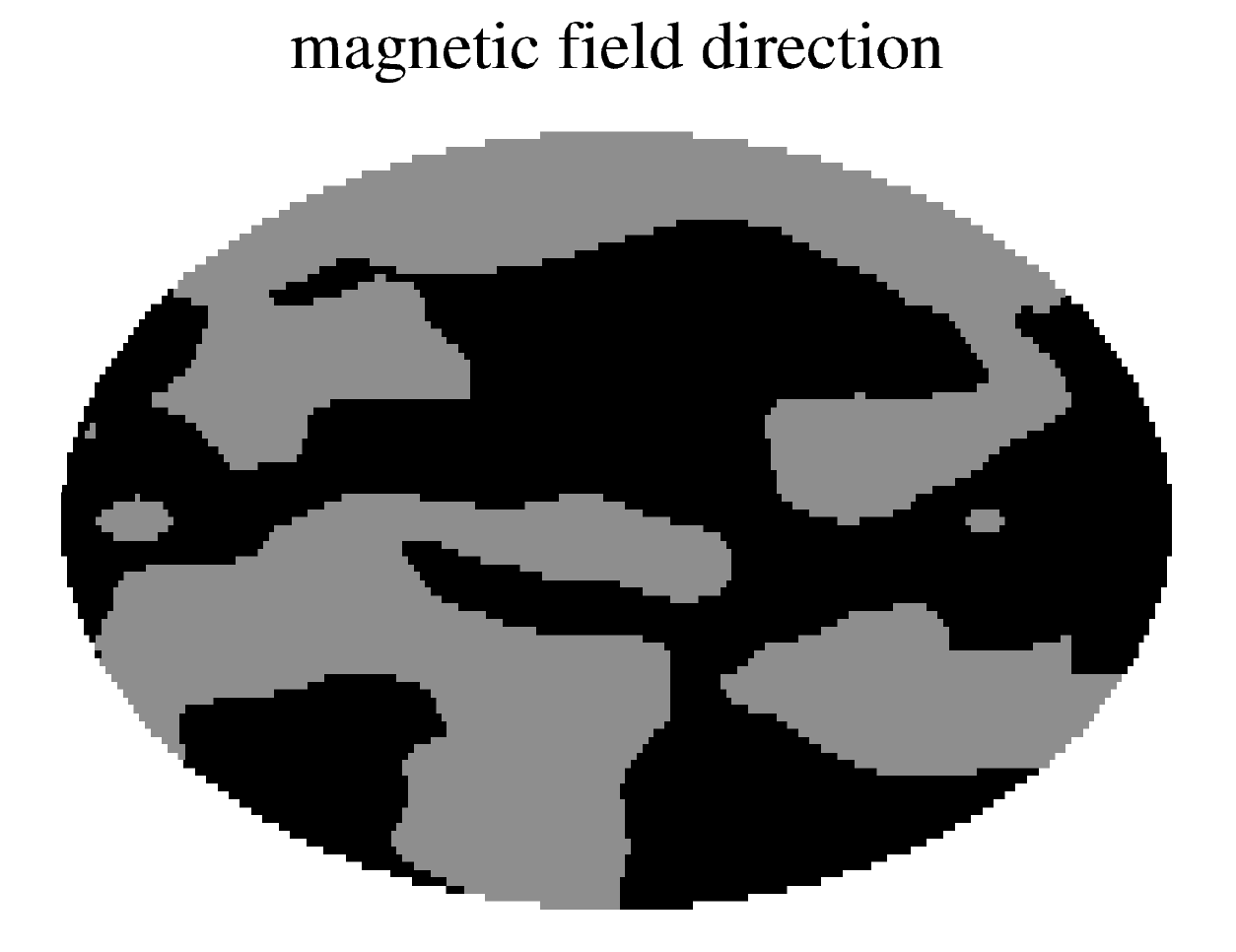} \\
 \includegraphics[width=0.24\linewidth]{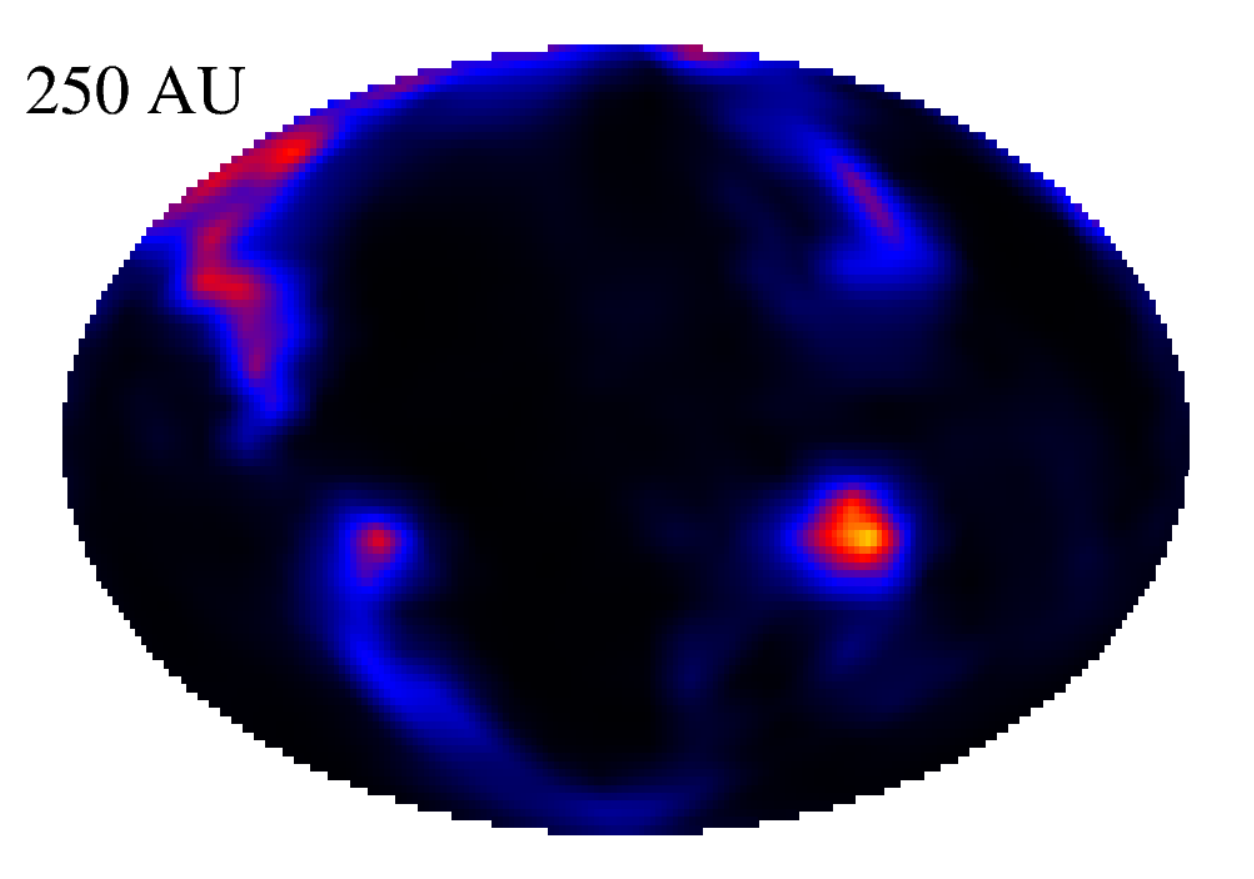}
 \includegraphics[width=0.24\linewidth]{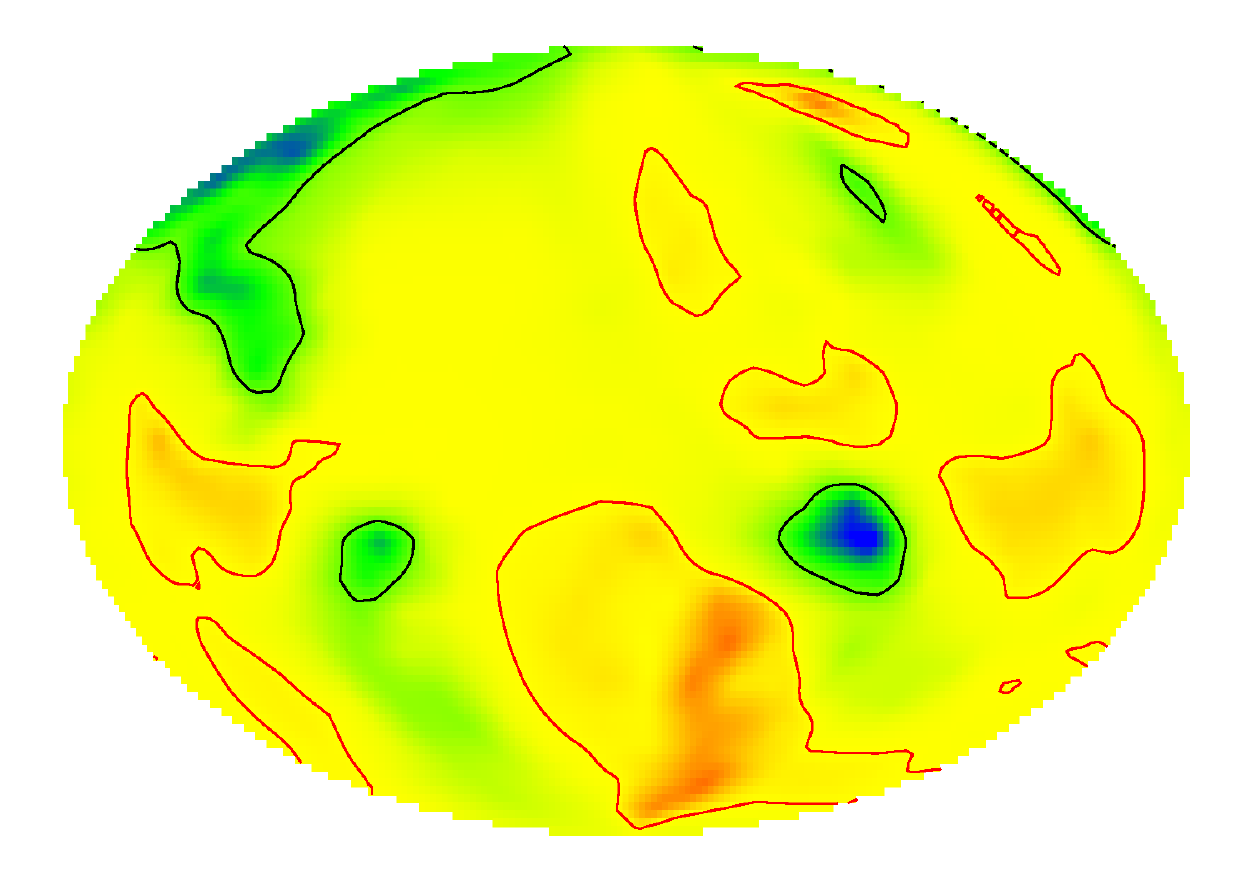}
 \includegraphics[width=0.24\linewidth]{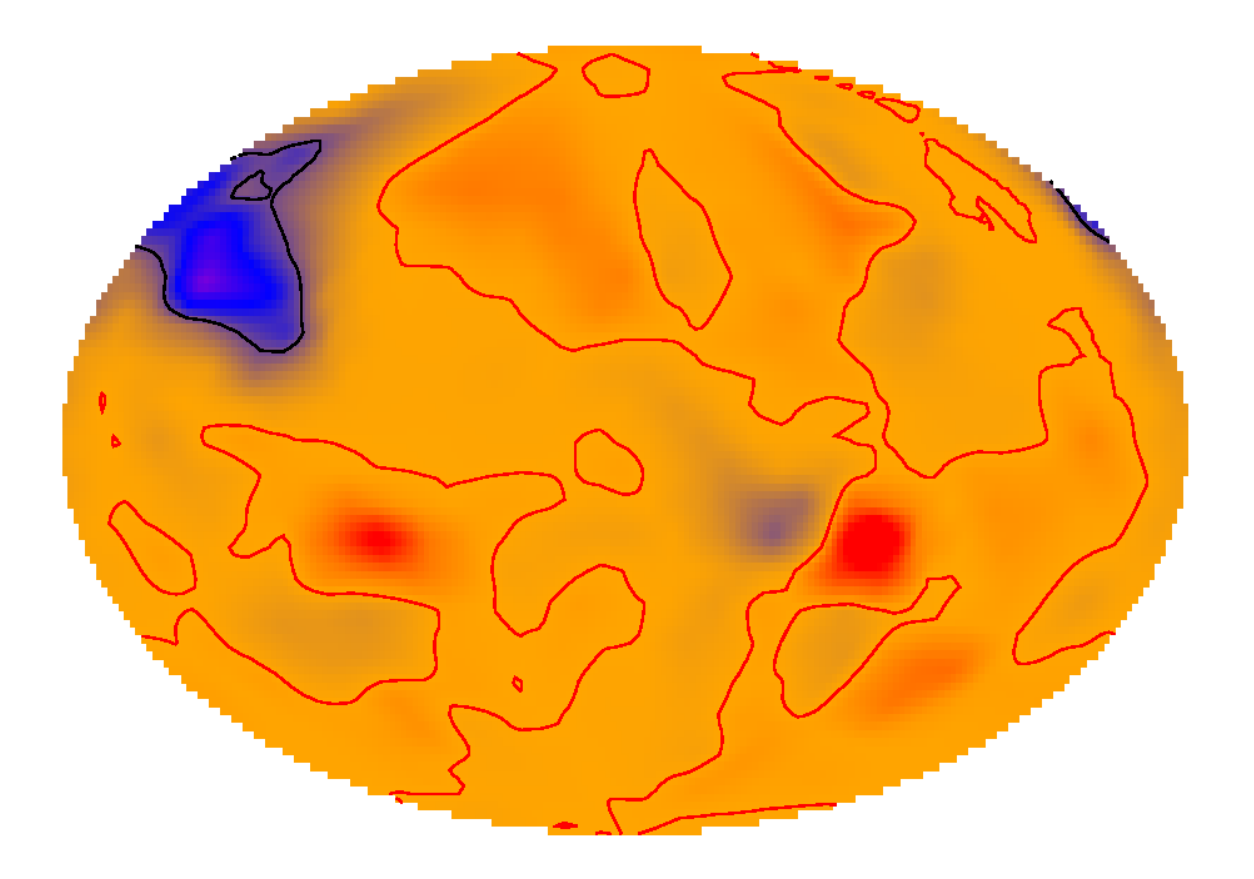}
 \includegraphics[width=0.24\linewidth]{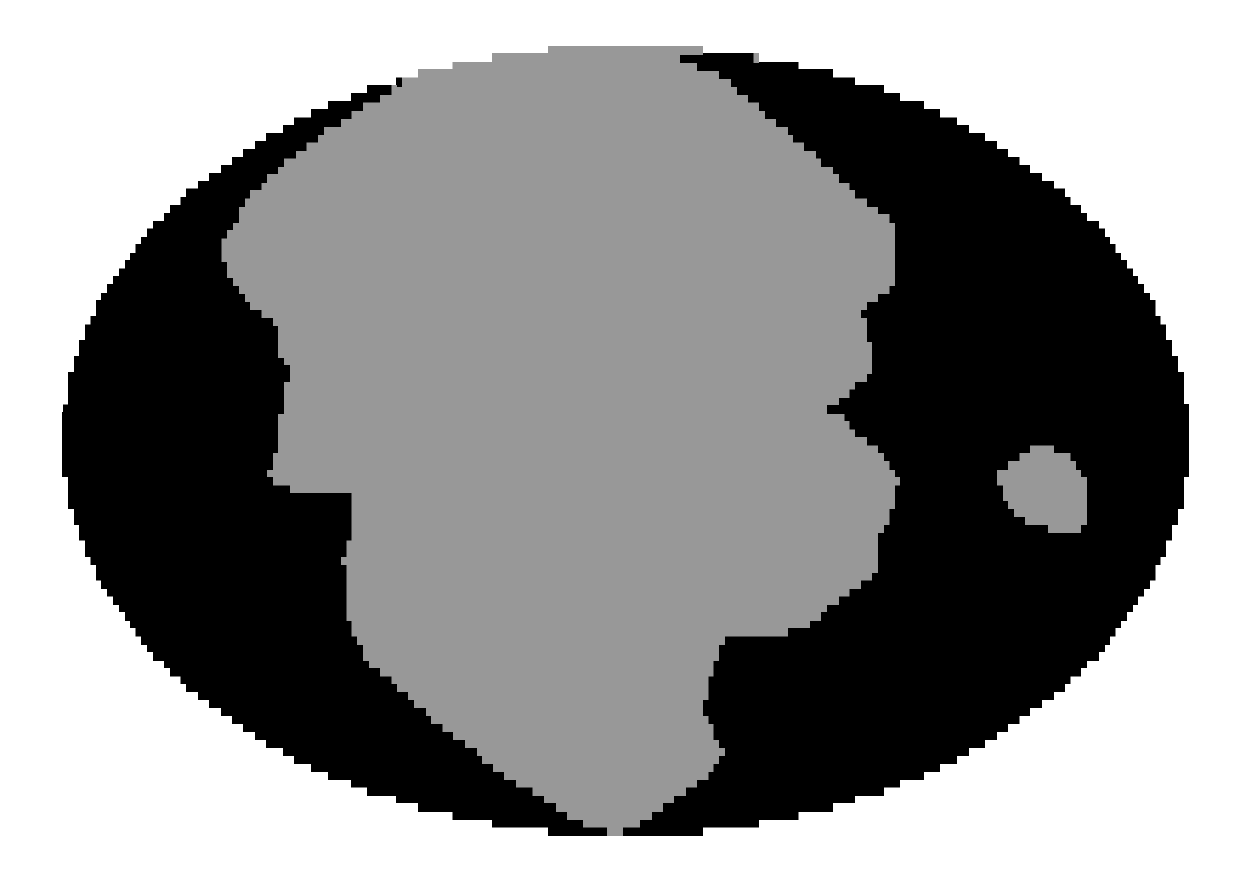} \\
 \includegraphics[width=0.24\linewidth]{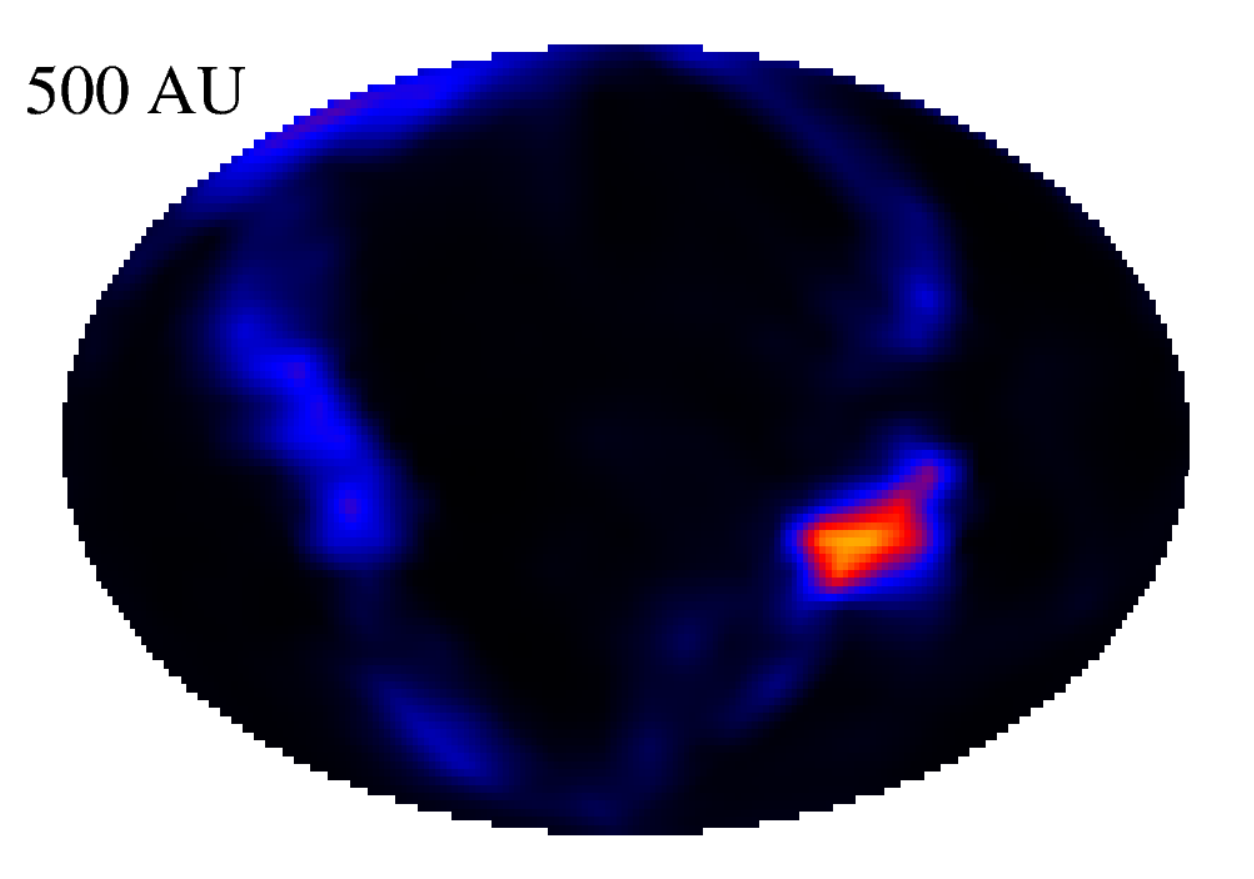}
 \includegraphics[width=0.24\linewidth]{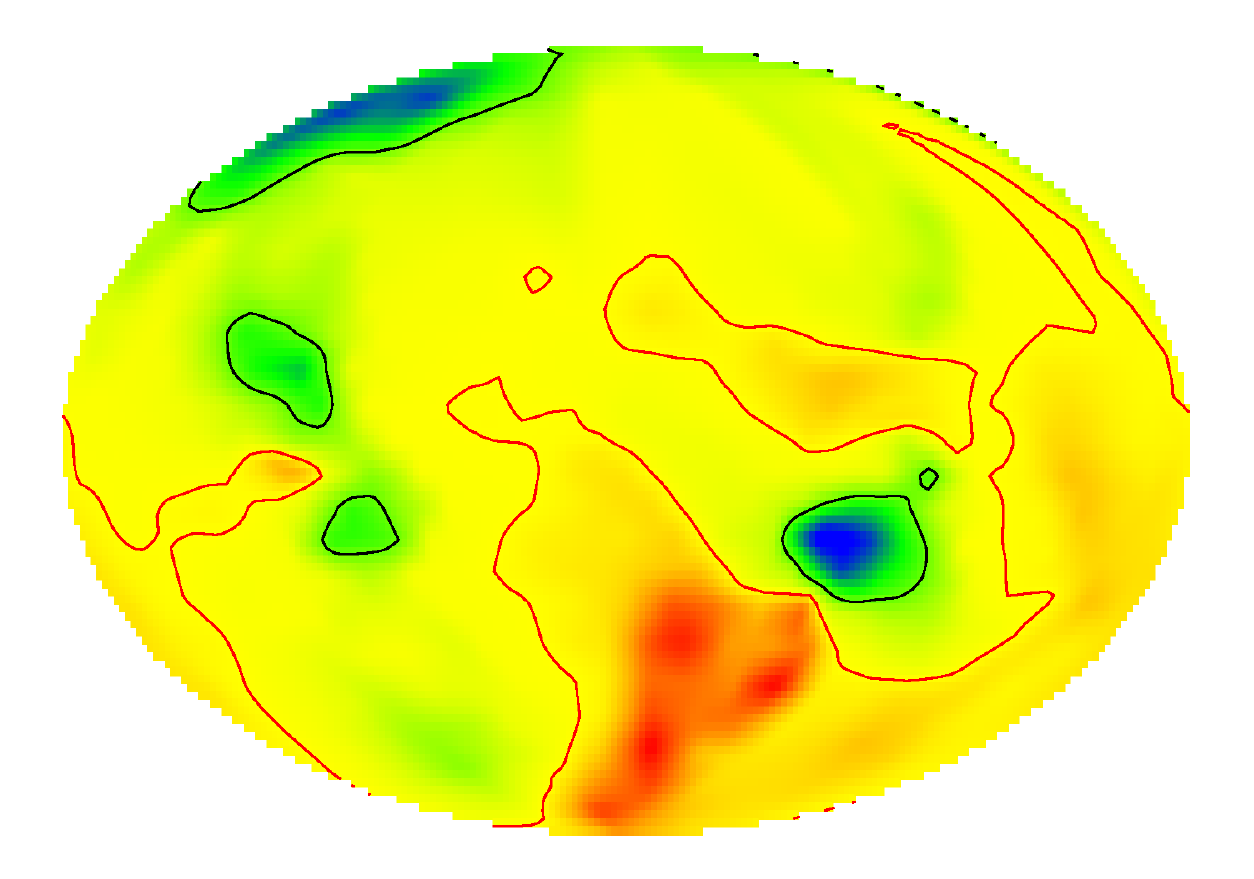}
 \includegraphics[width=0.24\linewidth]{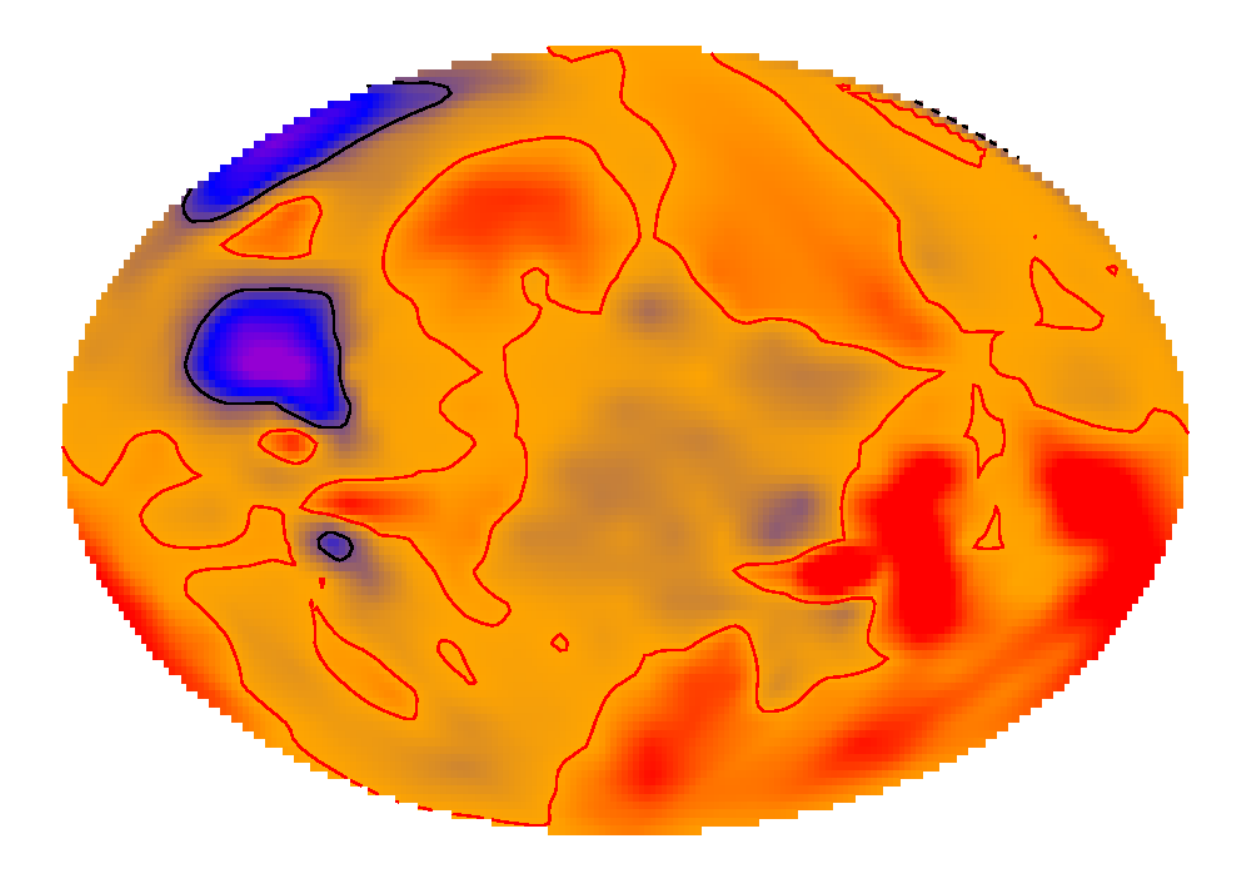}
 \includegraphics[width=0.24\linewidth]{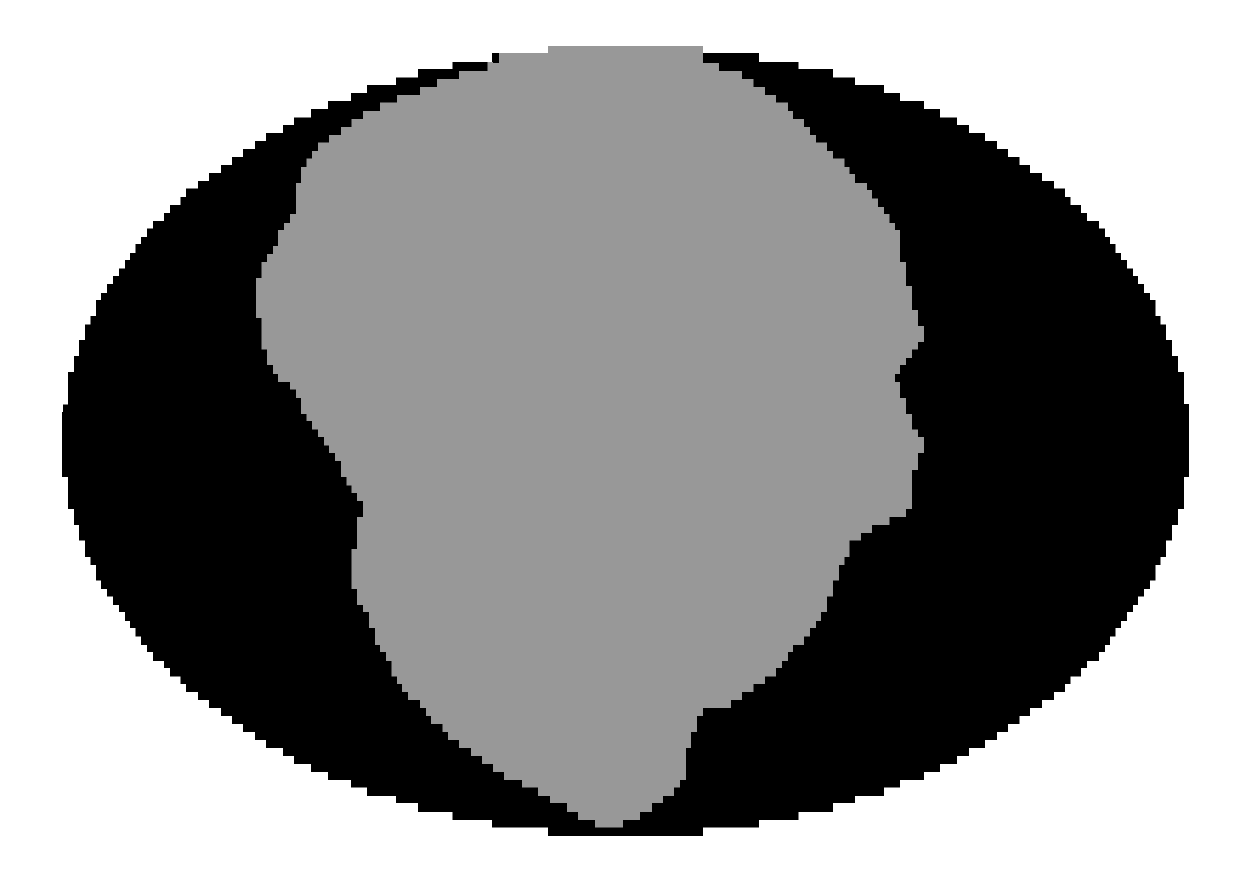} \\
 \includegraphics[width=0.24\linewidth]{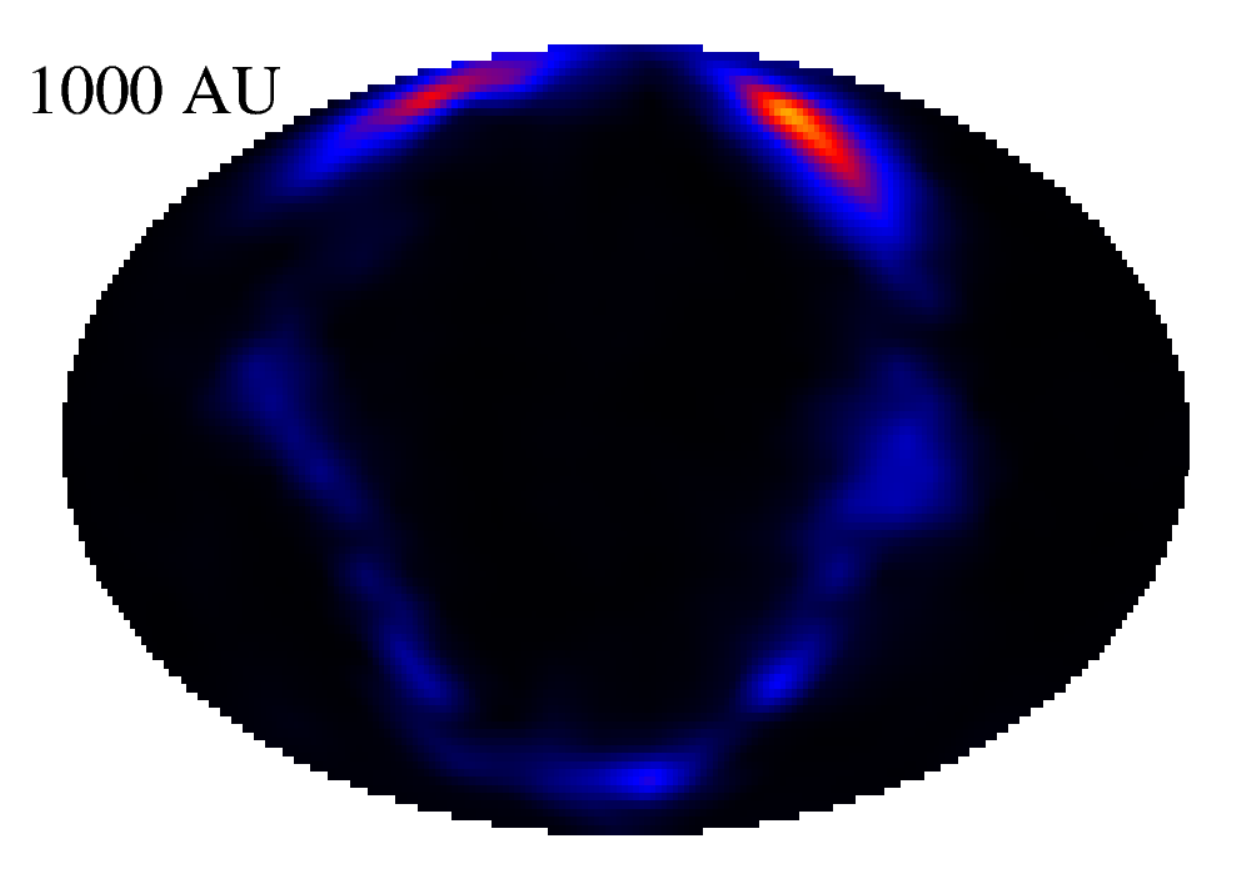}
 \includegraphics[width=0.24\linewidth]{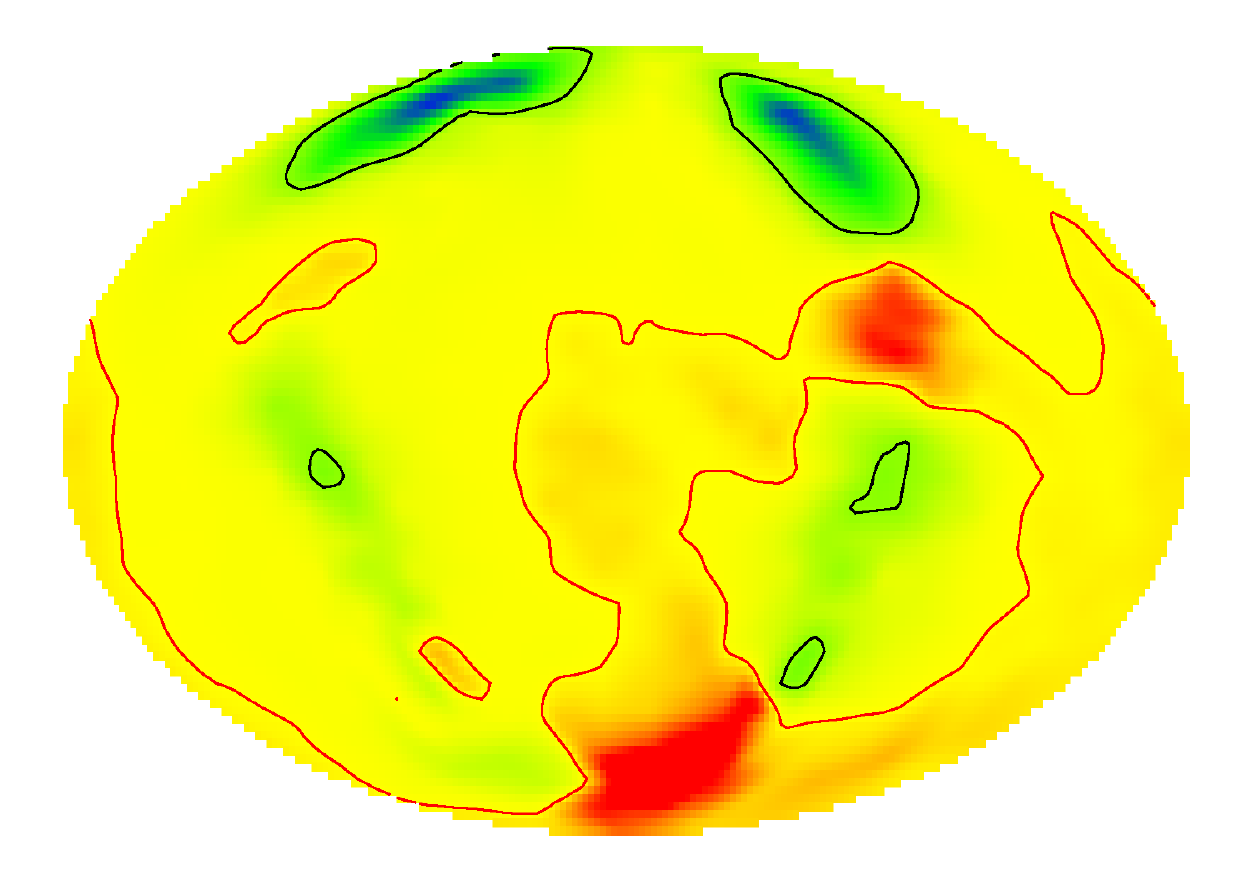}
 \includegraphics[width=0.24\linewidth]{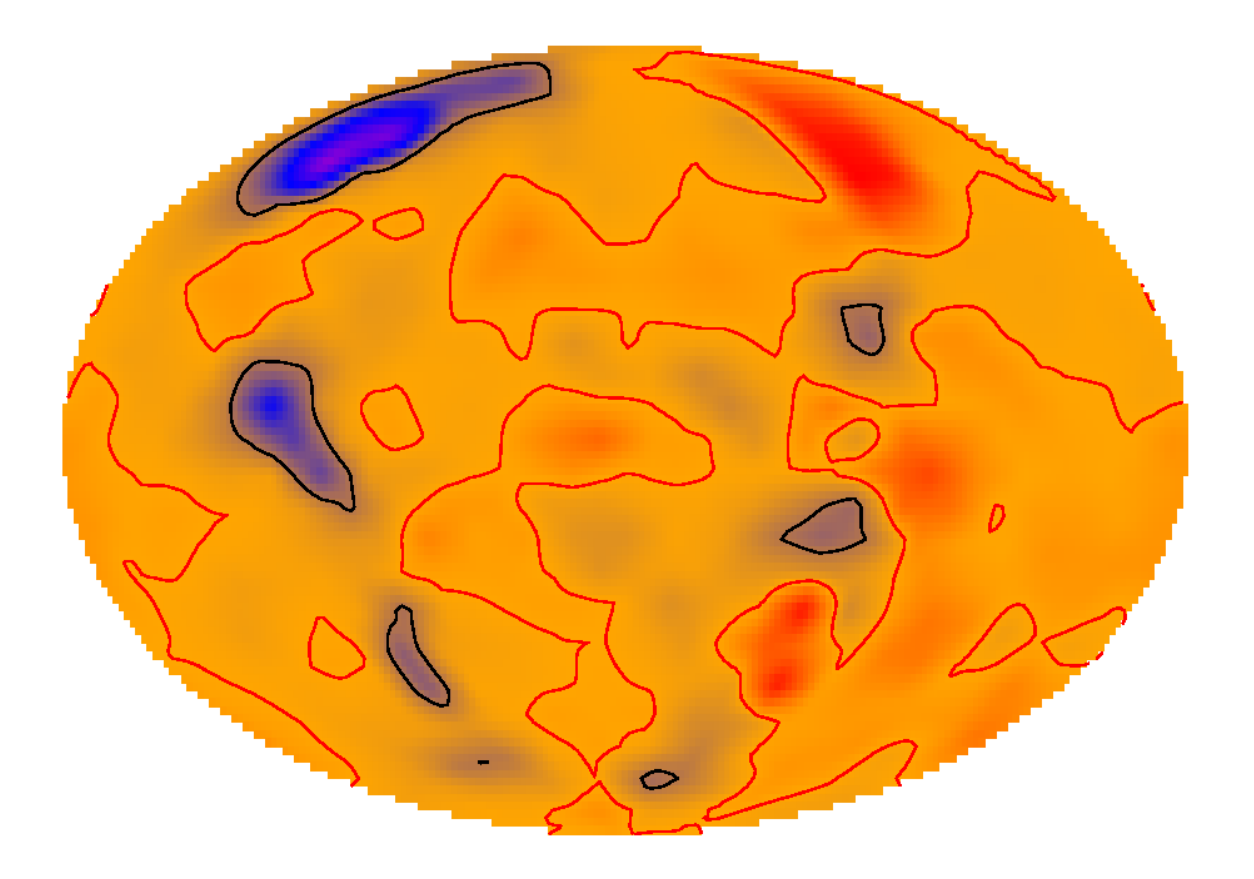}
 \includegraphics[width=0.24\linewidth]{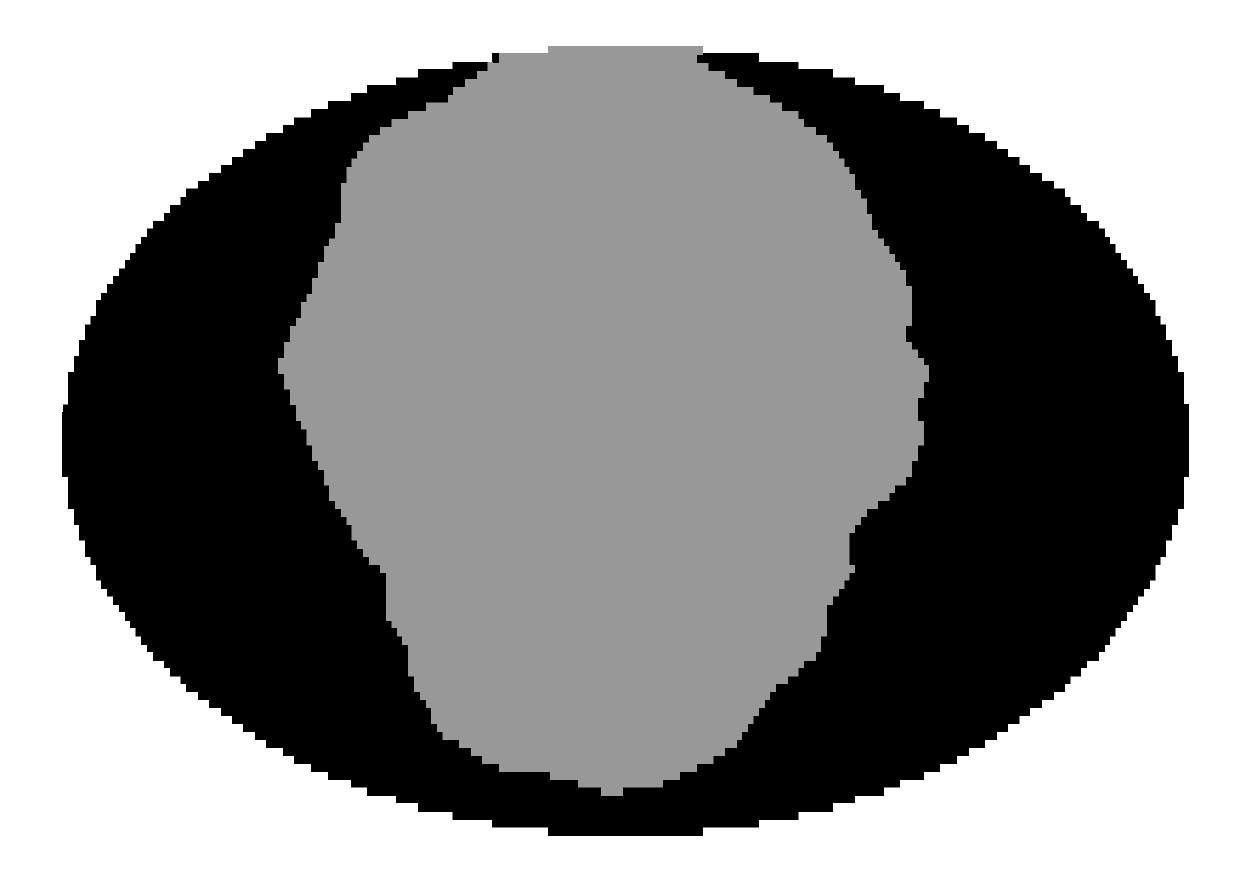} \\
 \caption{Left to right: Density, mass accretion flow, angular momentum accretion flow, and Hammer projection of the direction with which the magnetic field pierces through the surface of the spheres with radii of 50 AU, 250 AU, 500 AU, and 1000 AU (top to  bottom) around the protostellar disc at the end of run M2-NoRot. Second and third column: The red lines separate regions of accretion from regions of outflowing gas. Second column: The yellow-green-blue coloured regions represent mass flow towards the disc, red regions mass flow away from the disc. Third column: Angular momentum flow towards the disc occurs in the orange-blue regions. The highly anisotropic accretion structure is clearly recognisable in all panels. The regions enclosed by black lines are the calculated accretion channels (see text). Fourth column: Black denotes an ingoing direction of the magnetic field, grey an outgoing direction.}
\label{fig:accr}
\end{figure*}
We use a Hammer projection\footnote{A Hammer projection is a systematic transformation of locations on a sphere onto locations on a plane such that phenomena per unit area are shown in correct proportion (equal-area map).} in order to present the accretion/density structure at the surface of the sphere. Although the Hammer projection is not conformal, it is area conserving which is of particular importance for calculating the accretion rates through certain regions which is done below. In Fig.~\ref{fig:accr} we show the Hammer projection of the density as well as of the mass and angular momentum accretion flow for the distances of 50, 250, 500, and 1000 AU around the protostellar disc at the end of run M2-NoRot, hence at the same time as shown in Fig.~\ref{fig:overview}. As can be seen, none of the quantities reveals a well-ordered structure. In particular, the mass accretion flow (second column) clearly shows a highly anisotropic behaviour. A significant fraction of the accretion seems to occur through a low number of small, distinct areas with particularly high accretion rates well above the average (green-blue regions). Furthermore, also the entire region of accretion (separated by red lines from the regions where gas flows away from the disc), reveals a highly complex shape repeatedly intercepted by regions of outflowing material (red regions). We emphasise that for the sphere at 50 AU (top row of Fig.~\ref{fig:accr}) the effect of the protostellar disc is clearly recognisable for the density, mass and angular momentum accretion flow\footnote{We note that the sphere was rotated such that the disc's midplane coincides with the equator of the sphere.}. On larger scales, a ring like structure appears (most prominent in the density plots) which we attribute to the sheet formed due to the magnetic field-guided collapse in the midplane of the core perpendicular to the global field direction. Since the disc axis does not coincide with the global field direction, the ring structure is not located along the equator. We note that the anisotropic accretion flow connects smoothly to the results of \citet{Smith11} who also studied the accretion flow via Hammer projections although on a larger scales, i.e. from filaments onto molecular cloud cores. The authors showed that also on these scales accretion is far from spherically symmetric.

In order to further quantify the anisotropy of the accretion flow, we first determine the total accretion rate $\dot{M}_\rmn{tot}$ through the entire sphere. We note that for the calculation of $\dot{M}_\rmn{tot}$ we neglect the (negative) contribution of outflowing gas. Next, we define the integrated accretion rate $\dot{M}_\rmn{thres}$ as the accretion rate of all regions which have an accretion rate per area larger than some threshold $\dot{m}_{\rmn{thres}}$\footnote{We note that we use a capital $\dot{M}$ for the accretion rate integrated over the surface with units of g s$^{-1}$ and a small $\dot{m}$ for the accretion rate per area (g s$^{-1}$ cm$^{-2}$).}. We now adapt the threshold $\dot{m}_{\rmn{thres}}$ such that integrated accretion rate $\dot{M}_\rmn{thres}$ accounts for 50\% of the total accretion rate $\dot{M}_\rmn{tot}$. Given this threshold $\dot{m}_{\rmn{thres}}$, we can now analyse this area through which 50\% of the total accretion occurs, i.e. where $\dot{m}$ is larger than $\dot{m}_{\rmn{thres}}$. For anisotropic accretion this area is usually split up into several distinct regions (hereafter channels). Applying this analysis to run M2-NoRot, we can see that this is indeed the case for our simulations (second column of Fig.~\ref{fig:accr}). As can be seen, a few different channels enclosed by black lines are found, through which 50\% of the total mass accretion occurs.

Next, we determine the number of distinct channels as well as their integrated surface area A$_\rmn{channels}$, which we then compare to the entire surface area of all regions with accretion, A$_\rmn{accr}$. We first do this for the shell at 1000 AU (bottom panel in the second column of Fig.~\ref{fig:accr}). As already pointed out, the red line separates the regions of accretion (yellow-green-blue regions) from regions with outflowing gas. The total accretion area A$_\rmn{accr}$ accounts for about 70\% of the surface of the sphere. There are two large separated regions of accretion and three regions of outflowing material (obey the periodicity with respect to the left and right edge of the ellipsoid), which reveal a highly complex shape. This already indicates that accretion is far from being isotropic. The black lines enclose the channels which account for 50\% of the total accretion rate. As can be seen, there are 5 separate accretion channels. The integrated area of the channels A$_\rmn{channels}$ accounts for only 8\% of the total accretion area (which in turn is 70\% of the surface area). Hence, this demonstrates that in the case considered accretion is indeed dominated by a \textit{few narrow channels} -- a typical sign of anisotropic accretion.

We now extend this analysis to a large number of snapshots equally distributed over the entire time after the formation of the first sink particle. The results for the different spheres with radii of 50, 250, 500, and 1000 AU are given in Table~\ref{tab:accr} where we list the time-averaged number of accretion channels\footnote{Due to the time-averaging we also show the first decimal place for this value, despite the fact that the number of channels are integer values.} and the time-averaged ratio of the integrated area of channels to the area of the entire accretion region, A$_\rmn{channels}$/A$_\rmn{accr}$.
\begin{table*}
\centering
\caption{Results of the mass and angular momentum accretion flow analysis for run M2-NoRot for spheres with different radii. Listed are the time-averaged number of mass accretion channels and the ratio of the integrated area of the channels to the area of the entire accretion region (column two and three). Column four and five show the corresponding results for the analysis of the angular momentum accretion flow. The second-last column shows the overlap between mass and angular momentum accretion channels. The last column lists the mean number of magnetic field reversals.}
 \begin{tabular}{c|cc|cccc}
 Radius	&  \# channels	& A$_\rmn{channels}$/A$_\rmn{accr}$ & \# channels ($\bmath{L}$)	& A$_\rmn{channels}$/A$_\rmn{accr}$($\bmath{L})$ & Overlap & \# field reversals \\
\hline
  50	&  7.4		& 0.13	&  5.5	&  0.11		&  0.77 & 2.94 \\
 250	&  5.2		& 0.10  &  4.2	&  0.086	&  0.61 & 0.71 \\
 500	&  4.6		& 0.11  &  4.8	&  0.091	&  0.56 & 0.63 \\
1000	&  4.1		& 0.15  &  5.2	&  0.12		&  0.50 & 1.08 \\
\hline
 \end{tabular}
\label{tab:accr}
\end{table*}
As can be seen, even in the case of a weak (subsonic) initial turbulence field, the accretion flow in the surroundings of the protostellar disc is highly anisotropic during the entire evolution of the disc, occurring on average through about 4 -- 7 channels, which carry 50\% of the entire accreted mass. Although the mean number of channels decreases somewhat with increasing radius, accretion is not necessarily less chaotic/anisotropic. This is indicated by the low ratio A$_\rmn{channels}$/A$_\rmn{accr}$, which shows no clear trend with distance from the disc centre. Moreover, despite the fact that 50\% of the total accretion occurs through these accretion channels, on average they account for less than 15\% of the total area of accretion indicating very localised accretion flows. To summarise, throughout the evolution of the protostellar disc and up to distances of 1000 AU, accretion occurs preferentially through a few narrow channels indicative of a highly anisotropic accretion mode.

When applying this analysis to the angular momentum accretion flow, we find qualitatively and quantitatively very similar results (see column four and five in Table~\ref{tab:accr}). The bulk of angular momentum is again accreted through a few channels with a total cross section comparable to the one of the mass accretion channels. As can also be seen in Fig.~\ref{fig:accr}, there is in general a good spatial agreement between the mass and angular momentum accretion channels. A detailed calculation shows that on average about 50\% -- 70\% of the area of the angular momentum accretion channels overlap with that of the mass accretion channels (second last column of Table~\ref{tab:accr}).

\begin{figure*}
 \includegraphics[width=0.48\linewidth]{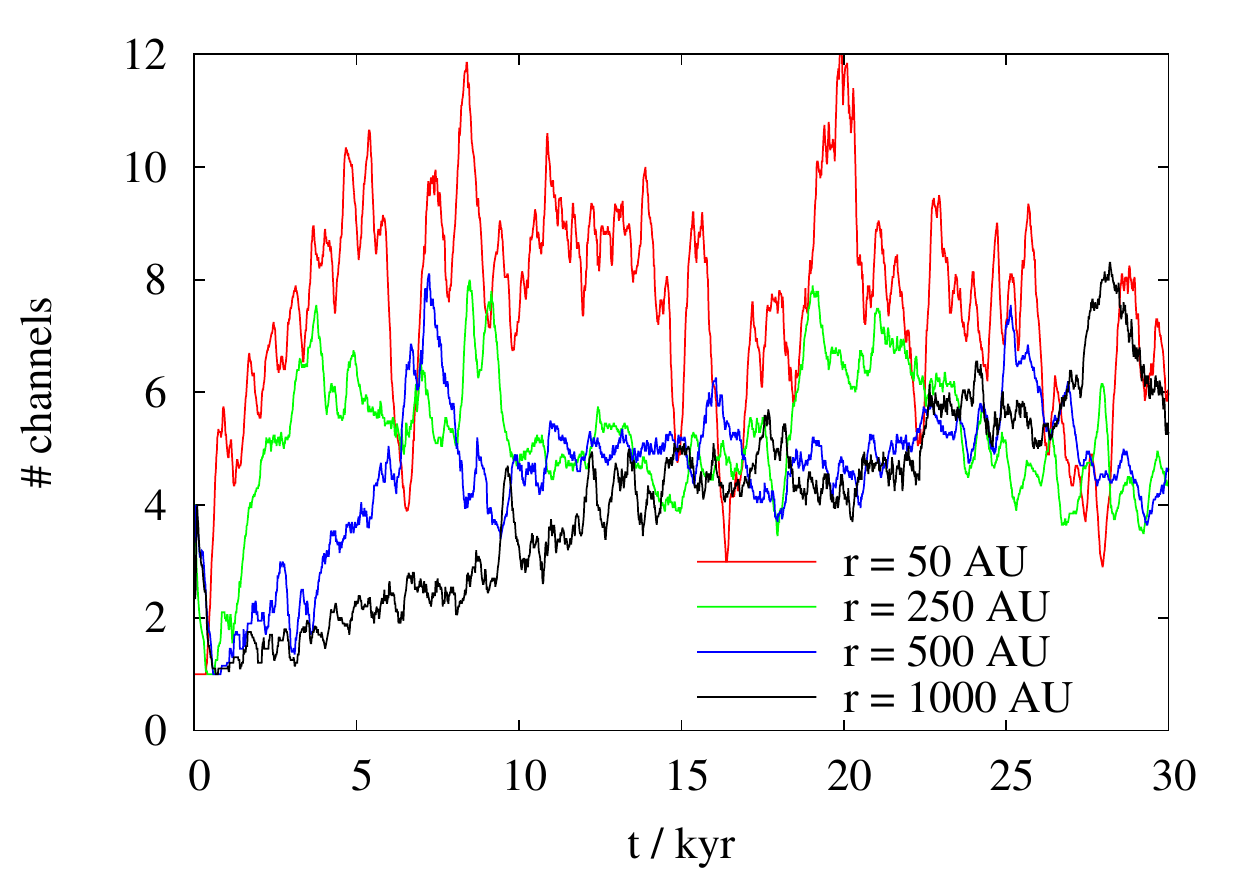}
 \includegraphics[width=0.48\linewidth]{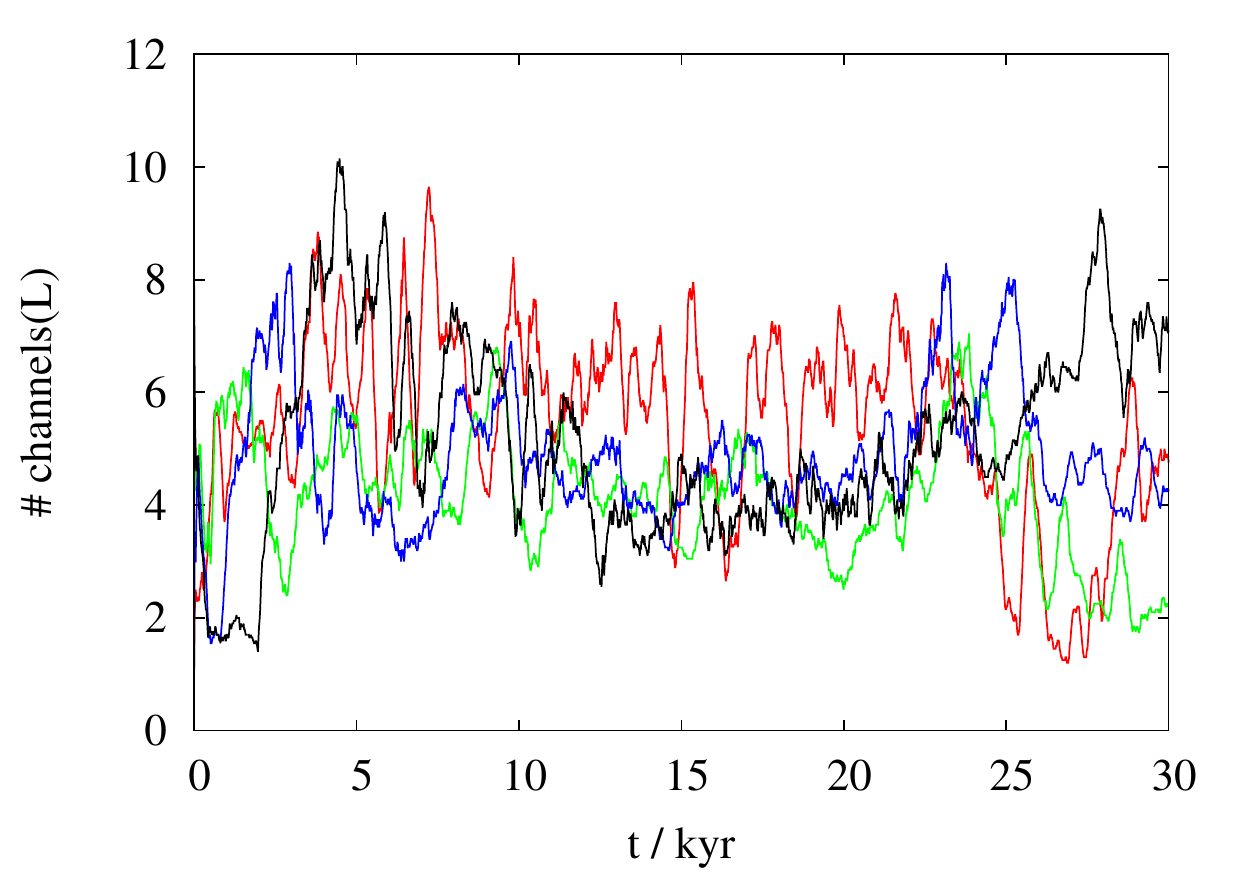}
\caption{Time evolution of the number of mass (left) and angular momentum (right) accretion channels for run M2-NoRot. In order to improve the readability of the plots, the numbers where smoothed over several timesteps.}
\label{fig:timeevol}
\end{figure*}

In Fig.~\ref{fig:timeevol} we show the time evolution of the number of mass and angular accretion channels for the four different radii considered before. We note that in order to improve the readability of the plots we averaged the number over a few dozen timestep, which is why non-integer values are present. As can be seen, the number of channels is subject to large variations in time, which we, however, rather attribute to the method for finding the channels than to real changes of the accretion structure: For two consecutive timesteps an area of high accretion could be described by a single channel or two channels depending on the actual value of the cutoff threshold. However, in general it can be seen that there is no systematic trend for the number of angular momentum accretion channels as well as that of the mass accretion channels at radii smaller than 500 AU. For a radius of 1000 AU, however, there is a slight increase over time recognisable in the number of mass accretion channels. This is caused by the ring like structure seen in Fig.~\ref{fig:accr}, i.e. the thin sheet in the midplane of the core, breaking up into smaller parts over time due to the impact of turbulence demonstrating its importance even at late times.

\subsubsection{Magnetic field structure}
\label{sec:fieldstructure}

\begin{figure}
 \includegraphics[height=0.48\linewidth]{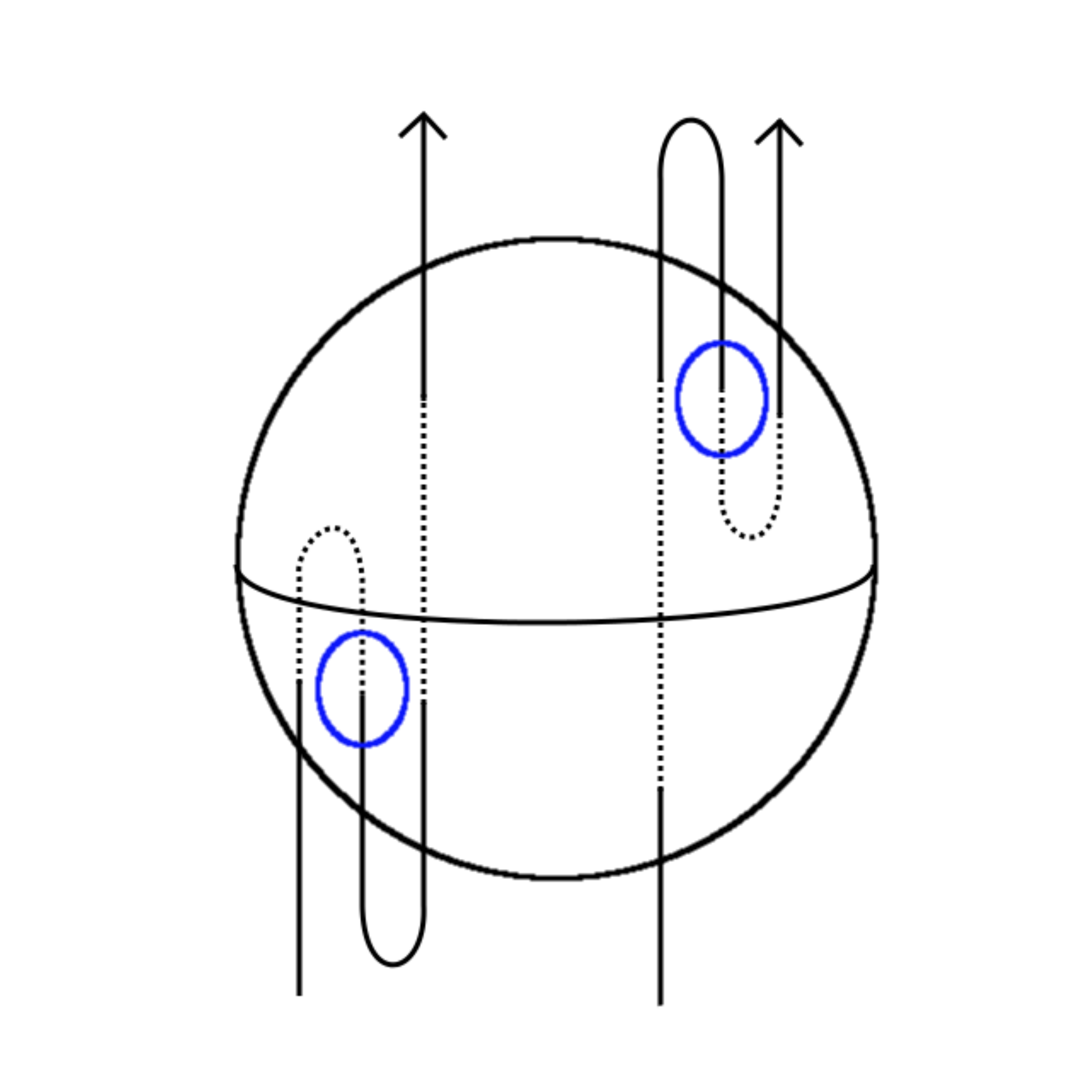}
 \includegraphics[height=0.42\linewidth]{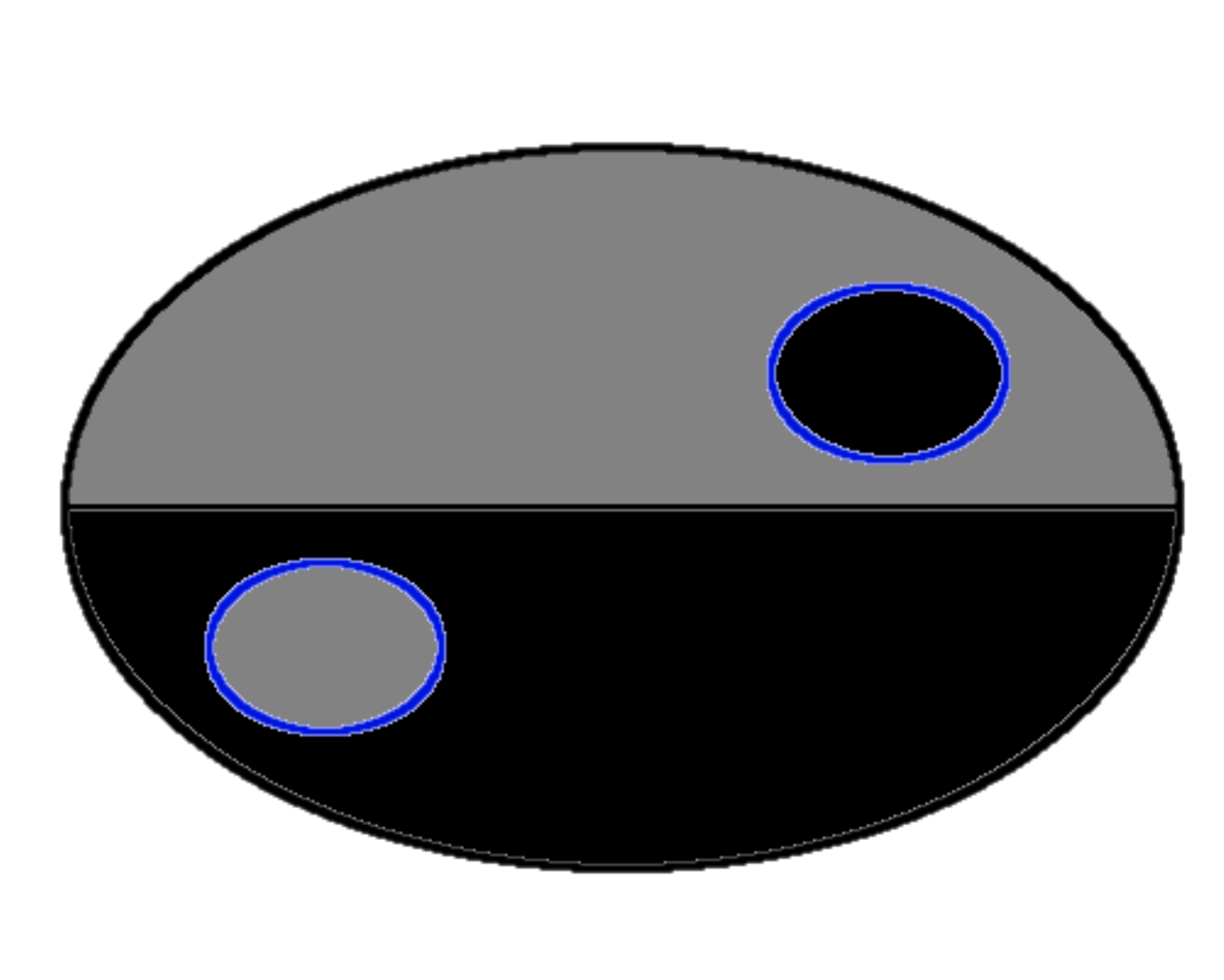}
\caption{Sketch of a magnetic field with two field reversals. Four distinct regions with different field directions at the sphere's surface are present. In the right panel the corresponding Hammer projection is shown. Grey denotes an outgoing direction of the field, black an ingoing direction.}
\label{fig:reversal}
\end{figure}

Next, we consider the direction with which the magnetic field pierces through the surface of the spheres with radii of 50, 250, 500, and 1000 AU around the protostellar disc in run M2-NoRot. We are only interested in whether the magnetic field points inwards or outwards, but not in the exact angle under which it pierces through the surface. In order to describe our approach for characterising the magnetic field structure in the vicinity of the discs, we first show a sketch of a magnetic field with two field reversals inside a given sphere (see Fig.~\ref{fig:reversal}).
The two field reversal result in two regions -- one in the upper and one in the lower half of the sphere's surface -- where the direction of the magnetic field piercing through the surface is flipped (enclosed by blue lines). We generate Hammer projections of the direction of the magnetic field in order to quantify field reversals at different radii. Analysing the projections we now determine the number of distinct regions in which the field pierces through the surface with the same direction. For the example shown in Fig.~\ref{fig:reversal} this would be two regions for \textit{each} direction. For any imaginable field configuration there have to be at least two distinct regions (one where the field goes in the other were it goes out). Hence, the \textit{minimum} number of field reversals inside the sphere (two in the example in Fig.~\ref{fig:reversal}) is simply the total number of distinct regions found (four in the example) minus two. We strongly emphasise that the real number of field reversals can actually be higher inside the sphere, which is why the number of field reversals inferred from the Hammer projections is only a \textit{lower limit} and thus also a lower limit for the degree of disorder of the magnetic field.

We first show the Hammer projections for the magnetic field direction of the spheres of with radii of 50, 250, 500, and 1000 AU around the protostellar disc at the end of run M2-NoRot (right column in Fig.~\ref{fig:accr}). We note that for a well-ordered, split monopole-like field configuration the Hammer projection would show two equally sized regions separated by the equator. On scales of 50 AU (top right panel) it can be seen that the magnetic field structure is more or less completely disordered. There are in total 6 distinct regions (5 for ingoing and 1 for outgoing magnetic fields), which means that there are \textit{at least} 4 field reversals inside the sphere\footnote{We remind the reader of the periodicity of the Hammer projection with respect to the left and right boundary of the ellipsoid.}. 

One can easily imagine that for this case the four field reversals are clearly a lower limit, since even the individual regions have a highly complex shape which points to an even more complex and disordered field inside the sphere. Considering the sphere with a radius of 250 AU (second row in Fig.~\ref{fig:accr}), the field appears to be somewhat better ordered since now only one field reversal is visible. Nevertheless, the structure of the largest regions is still quite complex, hence even here the field is far from being well-ordered. On even larger scales (500 AU and 1000 AU), however, the field becomes relatively ordered. We emphasise, however, that for other points in time not shown here field reversals do show up. However, the overall structure of the two regions for the large scales corresponds to that of a field which is more or less perpendicular the disc axis, i.e. it pierces through the sphere from ``left'' to ``right'' and not along the polar direction. This agrees with the sheet structure perpendicular to the field direction appearing as a ring like feature in the leftmost panels.

Next, we consider the time-averaged number of field reversals for the spheres with different radii around the protostellar disc (see last column of Table~\ref{tab:accr}). The results show that on scales of 50 AU there are on average several field reversals. Also on larger scales there is on average about one field reversal. Although this might seem low on first sight, we again emphasise that this analysis only gives a lower limit for the degree of complexity of the magnetic field. As noted before, a disordered field structure can reveal field reversals inside the considered sphere, which do not necessarily show up at the surface of the sphere. Furthermore, a single complex shaped region (see e.g. the black region in the top right panel of Fig.~\ref{fig:accr}) would point to a strongly tangled field structure despite giving only one field reversal in the method applied here. Hence, we conclude that despite the fact that at larger radii of up to 1000 AU only about 1 field reversal shows up in our analysis, even on these scales the magnetic field structure can be quite disordered.

In this context we emphasise that a relatively well-ordered, large-scale magnetic field inferred from low-resolution observations can be much more complex on small scales. This is also supported by recent observations \citep{Crutcher09,Crutcher10,Girart13} and numerical simulations \citep{Lunttila08,Bertram12}, which indicate that the magnetic field structure in molecular cloud cores is highly complex, possibly exhibiting several field reversals (see also Section~\ref{sec:discussion} for a more detailed discussion).

\subsection{High-mass core simulation}

\begin{figure*}
 \includegraphics[width=0.3\linewidth]{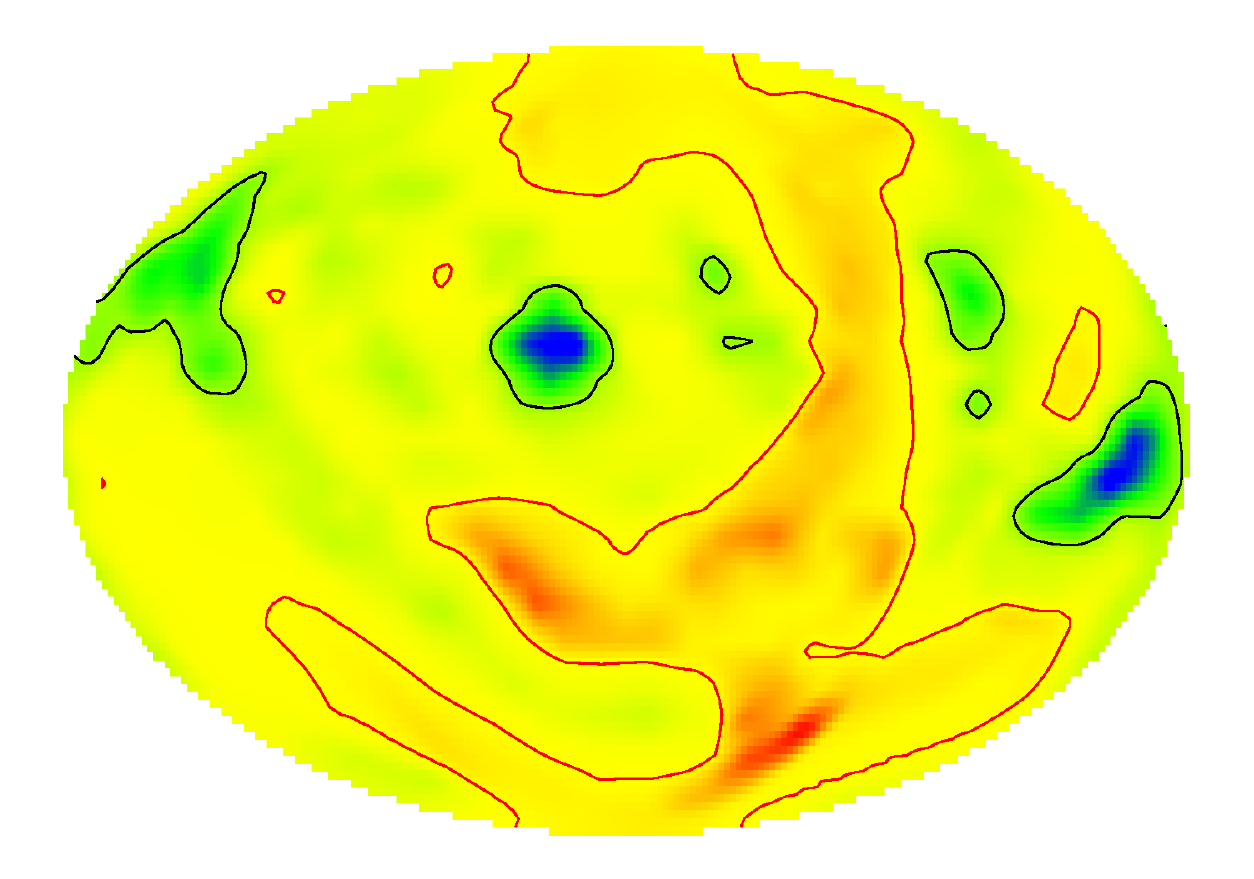}
 \includegraphics[width=0.3\linewidth]{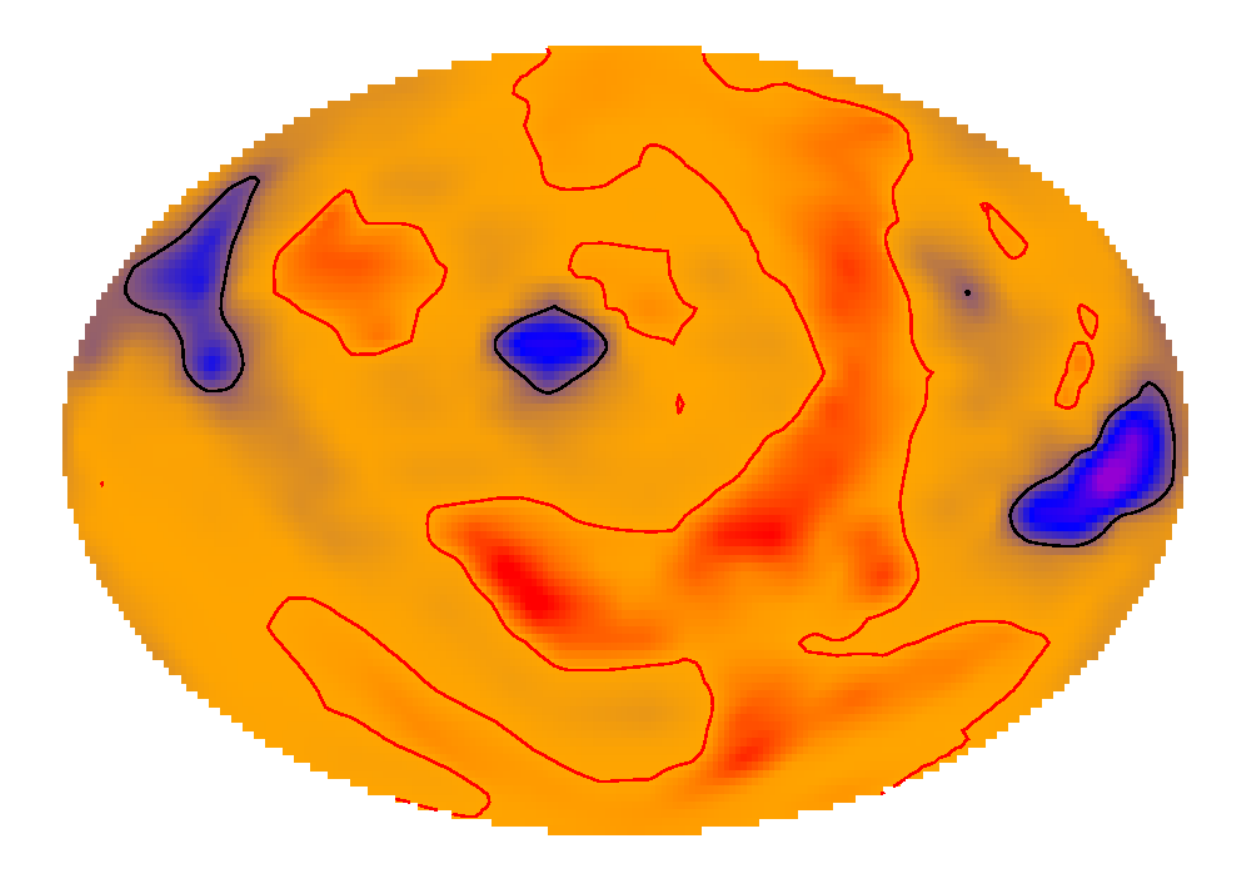}
 \includegraphics[width=0.3\linewidth]{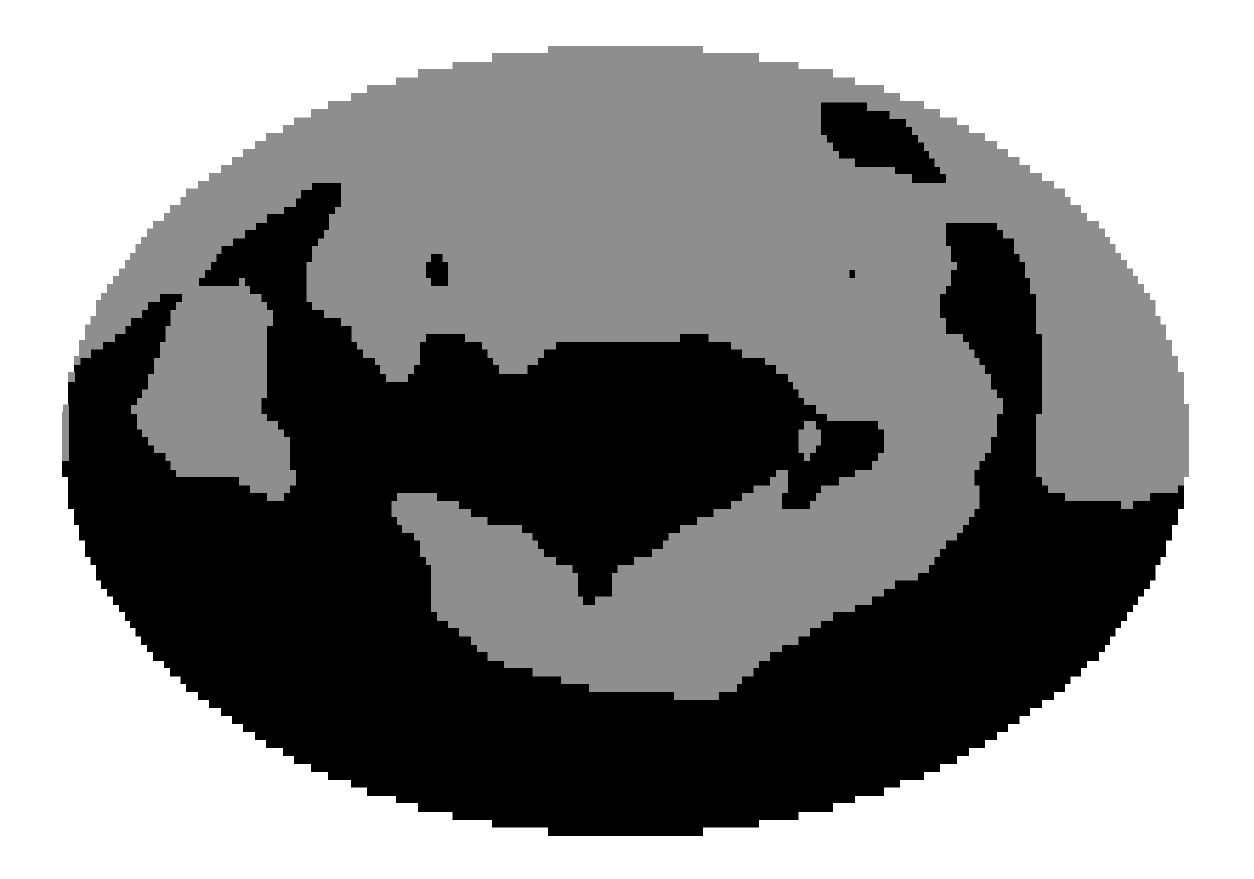}
 \caption{Mass accretion flow (left), angular momentum accretion flow (middle), and magnetic field direction through a sphere with a radius of 250 AU around the protostellar disc in run M100-NoRot after 25 kyr. Left and middle panel: The red line separates inflowing from outflowing material. Red regions represent mass and angular momentum flowing away from the disc.}
\label{fig:accrHM}
\end{figure*}

We now repeat the analysis done before for the high-mass, non-rotating core (run M100-NoRot). We first characterise the accretion flow around the protostar formed first in the simulation, which is also the most massive protostar having the largest associated protostellar disc. We first show the mass and angular momentum accretion flow through the surface of a sphere with a radius of 250 AU at the end of the simulation in Fig.~\ref{fig:accrHM}, the other radii of 50, 500, and 1000 AU are discussed quantitatively further below.
Despite the fact that accretion occurs through a large fraction of the surface of the sphere, it is again dominated by a few narrow channels randomly distributed over the surface.

In Table~\ref{tab:accrHM} we list the time-averaged number and size of the accretion channels calculated as described in the Section~\ref{sec:accrflow}.
\begin{table*}
\centering
\caption{Same as in Table~\ref{tab:accr} but for run M100-NoRot.}
 \begin{tabular}{ccccccc}
 Radius	& \# channels	& A$_\rmn{channels}$/A$_\rmn{accr}$ & \# channels ($\bmath{L}$)	& A$_\rmn{channels}$/A$_\rmn{accr}$($\bmath{L})$ & Overlap & \# field reversals \\
\hline
  50	&  6.2		&  0.103		&  4.4	&  0.090		&  0.74 & 3.43 \\
 250	&  6.6		&  0.073		&  4.4	&  0.061		&  0.69 & 1.31 \\
 500	&  7.0		&  0.075		&  4.2	&  0.060		&  0.71 & 1.16 \\
1000	&  7.1		&  0.086		&  4.3	&  0.072		&  0.69 & 1.18 \\
\hline
 \end{tabular}
\label{tab:accrHM}
\end{table*}
On average accretion of mass occurs along 6 -- 7 different channels. Although 50\% of the accretion is along these channels, they only account for about 10\% of the total area of accretion. Hence, these channels are indeed very localised and strongly pronounced (compare Fig.~\ref{fig:accrHM}). Just like for run M2-NoRot, accretion from the disc scale up to scales of 1000 AU appears to be highly anisotropic. Interestingly, there is no trend recognisable with varying radius, which indicates that the accretion flow is anisotropic on all scales considered. In this context we point out that since the disc extends over more than 50 AU in radius, the analysis for 50 AU refers to the accretion structure inside the disc. We interpret the appearance of anisotropic accretion even inside the disc as a consequence of the anisotropic accretion outside the disc as well as of the development of spiral arms which both result in more complex flow patterns. We also note that despite the much stronger initial turbulence the results shown in Table~\ref{tab:accrHM} do not differ significantly from that of run M2-NoRot (compare Table~\ref{tab:accr}). Analysing the flow of angular momentum reveals a similar picture without any clear trend with varying radius. Accretion of angular momentum seems to be even more constrained occurring through about 4 very narrow accretion channels (A$_\rmn{channels}$/A$_\rmn{accr} < 0.1$). Furthermore, as for run M2-NoRot the overlap between mass and angular momentum accretion channels is about 70\% thus in a reasonable range. Finally, we note that the time evolution of the number of mass and angular momentum accretion channels (not shown here) reveals a similar behaviour as that of run   M2-NoRot displayed in Fig.~\ref{fig:timeevol}.

Next, we analyse the magnetic field structure in the vicinity of the protostellar disc. As before we determine the number of field reversals at distances of 50, 250, 500, and 1000 AU by considering the direction with which the field pierces through the surface of the spheres. In the right panel of Fig.~\ref{fig:accrHM} we show the Hammer projection of the field direction at the surface of the sphere with radius of 250 AU around the most massive protostellar disc of run M100-NoRot. As can be seen, the magnetic field at that scale appears to be quite disordered exhibiting 4 field reversals.

Considering the averaged numbers of magnetic field reversals, we find that they are slightly increased compared to the numbers of run M2-NoRot (compare last column of Table~\ref{tab:accrHM} with that of Table~\ref{tab:accr}). This is a direct consequence of the stronger turbulent motions present in run M100-NoRot ($M_\rmn{rms}$ = 2.5 compared to 0.7 for run M2-NoRot). The disordered field structure remains recognisable up to scales of 1000 AU, where on average still more than one field reversal is present. Again, we emphasise that the number of field reversals listed in Table~\ref{tab:accrHM} is only a lower limit.

When considering the magnetic field direction at the position of the mass accretion channels for both the high- and low-mass run (see Fig.~\ref{fig:accr} and \ref{fig:accrHM}), it can be seen that the channels seem to be associated with a flip in the field direction. This is consistent with a picture of the magnetic field getting dragged inwards at the channel positions creating a 'cometary-like' field structure with the field lines bending backwards at the edges of the channels. We checked the importance of the magnetic field in accretion channels by comparing the magnetic field pressure $B/8 \pi$ with the ram pressure $\rho v_\rmn{radial}^2$ of the gas. We found that in general the ram pressure dominates over the magnetic pressure (as well as over the thermal pressure). Hence, it appears that the dynamics of the gas in these regions are dominated by the gravitationally driven collapse and not by the magnetic field. However, it is well possible that during the initial phase, when the accretion channels are shaped, the magnetic field might play a more important role. We note that a similar field configuration in regions of high mass accretion was also found by \citet{Li14}.

In summary, the low- and high-mass run show a highly anisotropic accretion flow structure on all scales as well as a highly complex, disordered magnetic field structure on small scales and -- in particular for the high-mass run M100-NoRot -- also on scales up to 1000 AU. The structure of the gas accretion flow and the magnetic field is therefore significantly different from the usually adopted structure of a split monopole (or homogeneous magnetic field) and a coherently rotating environment frequently used to assess the magnetic braking efficiency (see also the discussion in Section~\ref{sec:discussion}).

\section{Dependence on the initial conditions}
\label{sec:ICdependence}

In the following we extend our analysis to simulations with a wider range of initial conditions using the same methods as before.

\subsection{Dependence on turbulence strength}

Since we argue in \citet{Seifried12,Seifried13} that the turbulent motions are responsible for the formation Keplerian discs, it is a crucial task to check what minimum strength the turbulent motion must have in order to allow Keplerian discs to form. For the low-mass run M2-NoRot we have seen that already moderate subsonic turbulent motions are sufficient to enable disc formation. For the case of a high-mass molecular cloud core (run M100-NoRot) we have used a supersonic turbulence field with $M_\rmn{rms}$ = 2.5. Such supersonic turbulence is typical for high-mass cores \citep[e.g.][see also the review of~\citealt{Ward07} and references therein]{Caselli95,Andre07}. However, there might be massive cores with significantly weaker turbulent motions. In this case it is not clear a priori whether the formation of Keplerian discs would be possible since in massive, highly gravitationally unstable cores the dynamics are substantially different compared to low-mass cores. In particular the collapse and formation of the first protostar occurs on a much shorter timescale of about 20 kyr (run M100-NoRot) compared to $\sim$ 200 kyr in the low-mass core (run 2-NoRot).

For this reason we have performed three further simulations of 100 M$_{\sun}$ cores, M100-NoRot-Mrms0.5, M100-NoRot-Mrms1, and M100-NoRot-Mrms2.5, with turbulence strengths ranging from subsonic to supersonic values ($M_\rmn{rms}$ = 0.5, 1, and 2.5). We emphasise that run M100-NoRot-Mrms2.5 with $M_\rmn{rms}$ = 2.5 differs from run M100-NoRot since a different initial random seed for the turbulence field was used. Also for the other two runs discussed in this section different random seeds for the initial turbulence field are used. We follow the simulations over about 10 kyr after the formation of the first sink particle, which we suppose is sufficient since for corresponding runs discussed in \citet{Seifried12,Seifried13} Keplerian discs were already present after about 5 kyr.

\begin{figure*}
\centering
 \includegraphics[width=\linewidth]{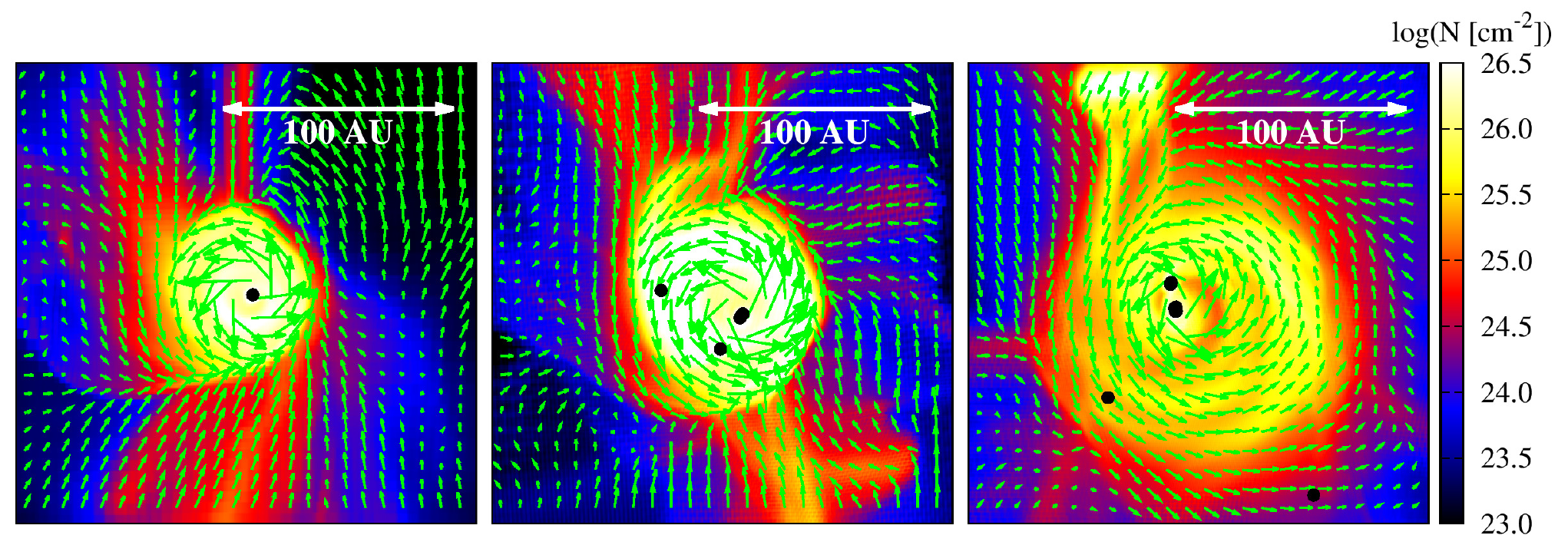}
 \caption{Column density and velocity vectors of the protostellar discs formed in the runs M100-NoRot-Mrms0.5, M100-NoRot-Mrms1, and M100-NoRot-Mrms2.5 (from left to right) seen from the top. The figures are 200 AU  in size.}
\label{fig:discsend}
\end{figure*}

We first consider the state of the discs at the end of each simulation in Fig.~\ref{fig:discsend}.
As can be seen, in all three runs a rotating structure has developed. By comparing the rotation velocity with the Keplerian velocity we found that the discs are indeed rotationally supported. Interestingly, the disc sizes decrease with decreasing amount of turbulent energy available in the core. The diameter of the discs are $\sim$ 65 AU in run M100-NoRot-Mrms0.5, $\sim$ 80 AU in run M100-NoRot-Mrms1, and about 120 AU in M100-NoRot-Mrms2.5. This indicates that there might by some general correlation between disc size and turbulence strength. For a thorough analysis of such a correlation, however, many more simulations would be required to exclude statistical effects, which by now is not feasible due to the high computational resources required. In summary, however, our simulation results clearly show that even for a relatively \textit{weak} turbulence field a protostellar disc can build up.

Analysing the accretion flow and the magnetic field structure in the surroundings of the three discs reveals similar results as for run M100-NoRot (see Table~\ref{tab:accrturb}). The accretion of mass and angular momentum is again highly anisotropic occurring along a few narrow accretion channels. Again, we point out that in particular for run M100-Rot-Mrms2.5 and partly for run M100-Rot-Mrms1 the analysis at 50 AU might refer to locations already inside the disc. As mentioned before, the similar number of accretion channels found at the radii of 50 AU and 250 AU might reflect the imprint of the anisotropic accretion flow at larger scales impacting on the discs. Interestingly, concerning the number and size of the accretion channels there is no clear trend recognisable for varying turbulence strengths. Similar also holds for the number of field reversals which varies between one and two for the different runs and distances considered. To summarise, even in massive molecular cloud cores with only a weak, subsonic turbulence field rotationally supported discs can build up surrounded by an anisotropic accretion flow and a complex-shaped magnetic field.
\begin{table*}
\centering
\caption{Same as in Table~\ref{tab:accr} but for the runs M100-NoRot-Mrms0.5, M100-NoRot-Mrms1, and M100-NoRot-Mrms2.5.}
 \begin{tabular}{cccccccc}
 Run  &  Radius	& \# channels	& A$_\rmn{channels}$/A$_\rmn{accr}$ & \# channels ($\bmath{L}$)	& A$_\rmn{channels}$/A$_\rmn{accr}$($\bmath{L})$ & Overlap & \# field reversals \\
\hline
M100-Rot-Mrms0.5 &  50	&  6.18	&  0.10		&  4.0	&  0.087		&  0.64		& 1.40  \\
& 250	&  5.5		&  0.066		&  3.8	&  0.056		&  0.58		& 0.64  \\
& 500	&  5.8		&  0.074		&  4.2	&  0.067		&  0.54		& 1.00  \\
& 1000	&  5.2		&  0.087		&  4.8	&  0.093		&  0.52		& 2.33  \\
\hline
M100-Rot-Mrms1 &  50	&  5.1	&  0.069	&  2.7	&  0.050		&  0.74		& 1.5  \\
& 250	&  5.8		&  0.057		&  2.7	&  0.039		&  0.72		& 0.71  \\
& 500	&  6.3		&  0.070		&  3.0	&  0.059		&  0.66		& 1.58  \\
& 1000	&  5.0		&  0.085		&  3.2	&  0.10			&  0.59		& 1.66  \\
\hline
M100-Rot-Mrms2.5 &  50	&  5.2	&  0.095	&  3.3	&  0.084		&  0.74		& 2.04  \\
& 250	&  4.9		&  0.070		&  2.3	&  0.054		&  0.76		& 0.71  \\
& 500	&  4.7		&  0.062		&  2.3	&  0.048		&  0.70		& 1.07  \\
& 1000	&  4.5		&  0.057		&  2.6	&  0.052		&  0.63		& 1.32  \\
\hline
 \end{tabular}
\label{tab:accrturb}
\end{table*}

\subsection{Accretion flow and magnetic field structure in the presence of global rotation}

It might be argued that the highly anisotropic accretion mode and the disordered magnetic field structure seen so far are simply a consequence of the lack of any global rotation in the simulated molecular cloud cores. For this reason we analyse two more runs M2-Rot and M100-Rot, which have identical initial conditions as run M2-NoRot and run M100-NoRot except an initial, moderate, global solid-body rotation of the entire cloud core superimposed on the turbulence field. The results of these runs were also discussed in \citet{Seifried13}. Just as for the other runs discussed before, a few 1000 yr after the formation of the first sink particle Keplerian discs are formed around them. In Table~\ref{tab:accrRot} we list the results of the accretion flow analysis for both runs M2-Rot and M100-Rot.
\begin{table*}
\centering
\caption{Same as in Table~\ref{tab:accr} but for runs M2-Rot and M100-Rot.}
 \begin{tabular}{cccccccc}
 Run  &  Radius	& \# channels	& A$_\rmn{channels}$/A$_\rmn{accr}$ & \# channels ($\bmath{L}$)	& A$_\rmn{channels}$/A$_\rmn{accr}$($\bmath{L})$ & Overlap & \# field reversals \\
\hline
M2-Rot &  50	&  5.0	&  0.13		&  4.6	&  0.11			&  0.72		& 3.71  \\
& 250	&  4.4		&  0.14		&  5.2	&  0.13			&  0.63		& 2.55  \\
& 500	&  4.3		&  0.16		&  5.7	&  0.15			&  0.63		& 1.73  \\
& 1000	&  4.9		&  0.22		&  6.7	&  0.18			&  0.65		& 2.74  \\
\hline
M100-Rot &  50	&  6.3	&  0.11		&  4.6	&  0.092		&  0.75		& 2.72  \\
& 250	&  6.8		&  0.067	&  4.2	&  0.054		&  0.73		& 0.77  \\
& 500	&  6.6		&  0.063	&  4.0	&  0.053		&  0.73		& 0.35  \\
& 1000	&  5.9		&  0.072	&  4.1	&  0.068		&  0.67		& 0.97 \\
\hline
 \end{tabular}
\label{tab:accrRot}
\end{table*}
As can be seen, even in the presence of global rotation the accretion of both mass and angular momentum reveals a highly anisotropic behaviour with accretion along a few, narrow channels. No clear trend for neither of the both runs is recognisable when varying the radius of the sphere. It appears that also in the presence of a global rotation the anisotropic accretion mode persists up to scales of 1000 AU. Interestingly, the magnetic field structure seems to be more disordered for the low-mass, weak turbulence run M2-Rot than for run M100-Rot. This, however, could change for different turbulence seed fields not tested here. As pointed out already, the number of field reversal, in particular the low value for 500 AU in run M100-Rot, might be an underestimate of the true degree of complexity of the magnetic field.

To summarise, even in the presence of an initial global rotation the accretion flow during the evolution of protostellar discs is far from being well-ordered and isotropic. Accretion of mass and angular momentum occurs along narrow channels. The magnetic field is strongly disordered with a few field reversals for both the low-mass and high-mass case.

\section{Discussion}
\label{sec:discussion}

\subsection{Mach number dependence}

As discussed in the introduction, in simulations of non-turbulent, collapsing cloud cores with magnetic field strengths comparable to observed strengths no rotationally supported protostellar discs were formed. This ``magnetic braking problem``, however, vanishes in the presence of turbulent motions \citep{Santos12,Santos13,Seifried12,Seifried13}.

A particularly interesting result of our present study is the fact that even a relatively low (subsonic) level of turbulence in massive prestellar cores can allow for the formation of Keplerian discs. As pointed out already before, in massive, highly gravitationally unstable cores the collapse proceeds much more rapidly. Hence, due to the higher infall velocities in such cores, turbulent motions of a fixed absolute strength (e.g. $M_\rmn{rms} = 0.5$) are less important compared to low-mass cores. This in turn means that no conclusions about disc formation in high-mass cores can be inferred from simulations of subsonic turbulence, \textit{low-mass} cores. However, given the fact that massive cores are usually observed to have at least transonic velocity dispersions ~\citep[e.g.][]{Caselli95,Andre07,Ward07}, our results strongly suggest that rotationally supported discs should be present in the Class 0 stage for most of the massive protostellar cores observed. This also holds for low-mass cores which usually have somewhat weaker subsonic turbulent motions. As shown earlier \citep{Santos12,Santos13,Joos13,Seifried13,Li14b}, and demonstrated in this work Keplerian discs are likely to form also in this case. Finally, we note that in our simulations we start with a initially smooth density distribution. Observed cores, however, always show some degree of clumpiness in their density distribution, which is why even for strongly subsonic turbulence we expect the development of non-spherical accretion flows. Taking these results together, we argue that already during the Class 0 stage rotationally supported discs are very likely to be formed for a wide range of protostellar masses.

\subsection{Magnetic flux loss}

It was argued by \citet{Santos12,Santos13} that the main reason for the formation of Keplerian discs in the presence of turbulence and strong magnetic fields is the reconnection of magnetic field lines aided by turbulence and the corresponding loss of magnetic flux. Despite the fact that we observe an increase of the mass-to-flux ratio $\mu$ in the vicinity of our discs \citep[see Fig. 7 in][]{Seifried13}, which at least partly can be attributed to an actual loss of magnetic flux \citep[see also][]{Li14b}, the importance of this effect for the formation of discs in the first place is not clear: To assess the efficiency of magnetic braking in the surroundings of the discs in such turbulence simulations, the mass-to-flux ratio in the region considered was frequently used. In a next step, non-turbulent collapse simulations with similar mass-to-flux ratios performed in the past \citep[e.g.][]{Allen03,Matsumoto04,Machida05,Banerjee06,Banerjee07,Price07,Hennebelle08,Hennebelle09,Duffin09,Commercon10,Seifried11} were considered. Based on the simulation outcomes, the efficiency of magnetic braking, one would expect, was extrapolated for the turbulence run. However, the non-turbulent simulations used to assess the magnetic braking efficiency reveal a well-ordered, large-scale magnetic field and a globally rotating core. As we could show, there is no such thing as global rotation or isotropic accretion\footnote{We note that strictly spoken in case of a non-turbulent, but magnetized collapse accretion is not isotropic either. However, in that case no variation with the azimuthal angle should be present.} nor a well-ordered magnetic field in the vicinity of protostellar discs as soon as sub- or supersonic turbulence is taken into account. Hence, such a comparison as described before should be considered with great caution.

Moreover, we point out that two effects contribute to the increase in $\mu$: On the one hand, there is magnetic diffusion, which might be enhanced due to the presence of turbulent motions, lowering the magnetic flux and increasing $\mu$ in the disc environment. On the other hand, accretion of mass unavoidably occurs partly along the field lines, which also contributes to the observed increase in $\mu$. This latter effect is not taken into account whenever calculating $\mu$ in a fixed volume possibly causing the importance of magnetic diffusion to be overestimated. We note, however, that from the numerical side it is hard to actually measure the amount of accretion along and perpendicular to the magnetic field lines and thus to estimate the relative importance of magnetic diffusion and accretion along field lines for the observed increase in $\mu$. Nevertheless, in order to do so, we calculated the accretion of mass along and perpendicular to the magnetic field in a sphere with a radius of 500 AU around the discs in the runs M2-NoRot and M100-NoRot. We find that of the order of 50\% of the observed increase in $\mu$ can actually be explained by accretion along field lines, i.e. without the need to consider the loss of flux due to magnetic diffusion. We note that the importance of mass accretion along field lines was recently also shown by \citet{Li14b}.

Furthermore, our results show a significant impact of turbulence on scales up to 1000 AU -- much larger than the grid scale where magnetic diffusivity is expected to act. Finally, as pointed out already in \citet[][see their figures 10 and 11]{Seifried11}, in non-turbulent simulations a large fraction of angular momentum is removed already \textit{before} the gas reaches the disc. Hence, a way to reduce the magnetic braking efficiency already before the gas falls onto the disc is required, which -- as shown here and confirmed recently by \citet{Li14b} -- can be simply achieved by the presence of turbulent motions.

\subsection{Comparison to observations}

Our findings also have implications for observationally based estimates of the likelihood that a rotationally supported disc can be formed: The low mass-to-flux ratio found in recent observations \citep[e.g.][]{Falgarone08,Girart09,Beuther10} does \textit{not} allow for the statement that Keplerian discs are unlikely to form since, as pointed out before, a disordered magnetic field structure with field reversals can significantly reduce the magnetic braking efficiency. Indeed, there are observations indicating that field reversals are present in molecular cloud cores: \citet{Crutcher09,Crutcher10} could show that in the outer parts of molecular cloud cores the line-of-sight component of magnetic fields partly has an opposite direction as that in the core centre. Moreover, the authors find that the mass-to-flux ratio in a good fraction of the observed cores seems to increase with increasing radius. This nicely fits with numerical simulations of \citet{Lunttila08} and \citet{Bertram12}, who have demonstrated that such a behaviour can be obtained through magnetic field reversals. Hence, these results indicate that the rather well-ordered magnetic field structure frequently observed in molecular cloud cores could be the result of an insufficient spatial resolution and that higher resolution observations are needed to unveil the true field structure. Indeed, observations of DR21 by \citet{Girart13} demonstrate how the telescope resolution affects the appearance of polarisation maps revealing an ever more complex and disordered field structure with increasing resolution. Taken together, these results -- although considering somewhat larger scales -- nicely complement ours, indicating the importance of turbulence for the structure of the magnetic field and thus the formation of protostellar discs.

\subsection{Misaligned magnetic fields}

Recently, it was shown that a magnetic field inclined with respect to the rotation of the core can result in the formation of rotationally supported discs \citep{Hennebelle09,Ciardi10,Joos12}. Based on these simulations and observational results, \citet{Krumholz13} showed, however, that this would result in a fraction of Keplerian discs in the Class 0 stage of about 10\% (at the very most 50\%). Given the fact that \textit{all} our simulations show Keplerian discs around Class 0 type protostars, we conclude that once turbulence is taken into account this fraction is pushed towards significantly higher values well above 50\% -- possibly\footnote{being aware of the low number statistics} even close to unity. We emphasise that this is in good agreement with disc fractions around Class I/II objects \citep[e.g.][but see also \citealt{Williams11} for a recent review]{Haisch01,Jorgensen09}. We note that \citet{Hull13,Hull14} (but see also \citet{Davidson11} and \citet{Chapman13} for contrary results) have recently shown that magnetic fields around protostars are partly strongly misaligned with the outflow/disc-axis. However, as already shown in \citet{Seifried12,Seifried13} (figure 3 and 7, respectively), the mean direction of a strongly tangled magnetic field with field reversals occurring up to scales of 1000 AU can be strongly inclined to the disc axis, thus in basic agreement with the results of the aforementioned authors. Hence, we argue that the magnetic field inclination found by \citet{Hull13,Hull14} might simply be an effect of a spatially unresolved, highly disordered magnetic field which supports our picture drawn in this work. This could also explain the apparent contradiction to the results of \citet{Davidson11} and \citet{Chapman13}, which give evidence for a preferentially aligned magnetic field. Synthetic polarisation maps of our simulations could shed light on how the telescope resolution affects the observed field structure -- a task we plan to tackle with a newly developed polarisation code \citep{Reissl14}. To summarise, we argue that in a turbulent environment the direction of $\langle\bmath{B}\rangle$ with respect to the disc/outflow-axis could be misleading when trying to estimate the braking efficiency.

\subsection{Open questions and uncertainties}

The anisotropic accretion of mass and angular momentum raises the question how in such a flow a rotating structure can be build up and in particular maintained. This can be understood by considering the position of the accretion channels. Since the discs reside in large-scale filamentary structures, which do not reveal significant positional changes during the time considered, also the orientation of the accretion channels relative to the disc do not exhibit excessively pronounced changes over time. Hence, despite being highly anisotropic, the time variability of the accretion flow is relatively moderate and results in a rather orderly build-up of the protostellar discs. In this context it would be interesting to see how a possible (pseudo-) disc warping might affect the transport/removal of angular momentum as suggested by \citet{Li14b}.

Recently, \citet{Machida14} have shown that Ohmic dissipation can lead to the formation of Keplerian discs for non-turbulent simulations which appears to be in contradiction with previous results \citep[e.g.][]{Krasnopolsky10,Li11,Dapp12}. In the context of this work, we would like to point out that \citet{Machida14} find an extremely high resolution ($<$ 1 AU) to be required to form a Keplerian disc. This demonstrates that the crucial effect responsible for disc formation in their simulations acts on relatively small scales. 
However, as we have pointed out before, our results show that other mechanisms are acting already on larger scales and allow for the formation of extended Keplerian discs. Most likely both processes are present at the same time but influence different spatial scales.

We again point out that our method to determine the number of field reversals can give us only a lower limit. In contrast to that the naive approach of calculating the mean of the magnetic field in a certain region and comparing it to the root-mean-square (as a measure of the fluctuations) is not sufficient. This can be understood by considering a purely toroidal field -- as it would be present in a disc -- which has a mean of zero but a non-vanishing rms value. This would result in a wrong assessment of the strength of the fluctuations and thus the degree of disorder. We therefore prefer to use the approach of measuring the number of field reversals despite the limitation that it can give only lower limits.

We note that the anisotropic accretion is also accompanied by an unsteady, episodic accretion onto the disc and subsequently onto the protostar itself. This could result in varying protostellar luminosities, which in turn was shown to have an effect on the properties and stability of protostellar discs \citep{Stamatellos12}. A detailed analysis of episodic accretion and varying luminosities, however, is deferred to a subsequent paper.

Finally we emphasise that our work nicely complements the recent work of \citet{Li14b} who demonstrate the importance of turbulence for effects like pseudodisc warping and magnetic reconnection. We argue that the impact of turbulence on scales of 1000 AU (and possibly even larger) as shown in our work, would promote the mechanisms for disc formation suggested by \citet{Li14b} in an even stronger way than probed by their highly idealised simulations. Hence, combining the work on disc formation found in literature, a more global picture seems to emerge: On larger scales turbulence seems to set the stage required for disc formation, i.e. providing a highly anisotropic accretion mode and a disordered magnetic field structure. These conditions, in turn, seem to be the prerequisites for the turbulence-enhanced loss of magnetic flux on smaller scales, e.g. by magnetic reconnection. In summary it appears that a combination of different effects -- on the disc scale as well as on significantly larger scales -- seems to contribute to the formation of rotationally supported protostellar discs.

\section{Conclusions}
\label{sec:conclusions}

We have performed simulations of turbulent, collapsing molecular cloud cores covering a wide range of initial conditions concerning core masses, turbulence strengths and the presence/absence of global rotation. We find that for all of our simulations rotationally supported discs build up within a few kyrs after the first protostar has formed. In particular, we showed that even for mild, i.e. subsonic turbulent motions the magnetic braking efficiency is reduced enough to allow Keplerian discs to form. Given this fact we suggest that the fraction of Keplerian discs around Class 0 stage protostars is significantly higher than 50\% -- likely even close to unity. This value is higher than the uppermost limit from simulations with magnetic fields misaligned with the core's angular momentum vector \citep{Krumholz13}. Furthermore, our result is consistent with to the observed disc fraction around Class I/II objects \citep[e.g.][]{Williams11}. It therefore demonstrates the importance of turbulent motions for the formation of Keplerian discs at very early stages.

We showed that under more realistic conditions, i.e. as soon as even mild ($M_\rmn{rms} \geq 0.5$) turbulent motions are included in simulations, the classical disc formation scenario of a coherently rotating environment and a well-ordered magnetic field breaks down: Analysing the surroundings of the protostellar discs in our simulations, we find that the accretion of mass and angular momentum occurs in a highly anisotropic manner out to distances of 1000 AU from the  discs. It is dominated by a few narrow accretion channels which carry a significant part of the inflowing mass and angular momentum but cover only about 10\% of the surface area. We find that the number and size of the accretion channels show only slight systematic trends over time. Inside these accretion channels, in which the ram pressure dominates over thermal and magnetic pressure, the magnetic field is dragged inwards towards the disc. This anisotropic accretion mode is found in all of our simulations, i.e. independent of the core mass, the turbulence strength, and the presence/absence of global rotation. Also within the discs accretion reveals an anisotropic behaviour which could be due to the larger-scale anisotropy as well as gravitational instabilities like spiral arms developing over time. Furthermore, we find that the magnetic field structure in the vicinity of the protostellar discs is highly disordered revealing field reversals on all scales considered. This demonstrates that the recently observed inclination of magnetic fields with respect to the disc axis \citep{Chapman13,Hull13,Hull14} could simply be the consequence of a spatially unresolved, highly disordered field structure.

To summarise, our simulations present a richer picture of the process of disc formation, wherein turbulence, filamentary accretion streams, and magnetic field reversals guarantee that the otherwise overwhelming strength of magnetic braking by ordered fields is significantly degraded, allowing Keplerian discs to form. 
Non-turbulent collapse simulations, on the other hand, might significantly overestimate the efficiency of magnetic braking and thus underestimate the fraction of Class 0 stage Keplerian discs. 
We suggest that the anisotropic accretion and disordered magnetic field structure found in the environment of protostellar discs might set the stage for other mechanisms contributing to the formation of discs like (pseudo-) disc warping or non-ideal MHD effects \citep[e.g.][]{Li14b,Machida14}. Finally, in order to improve our understanding of the process of protostellar disc formation, we suggest that further high-resolution observations unveiling the magnetic field structure and the dynamics in the direct vicinity of protostellar discs, e.g. by ALMA, are required. In this context, using the method presented in \citet{Reissl14}, we plan to produce synthetic polarisation maps of the simulations presented here, which would provide a valuable tool to interpret such observations.

\section*{Acknowledgements}

The authors like to thank the referee S. Van Loo for his constructive report which helped to significantly improve the paper. We also like to thank Zhi-Yun Li for helpful comments and discussions.
D.S. and R.B. acknowledges funding by the Deutsche Forschungsgemeinschaft via grants BA 3706/3-1 within the SPP \textit{The interstellar medium}, as well as via the Emmy-Noether grant BA 3706/1-1. 
D.S. further acknowledges funding by the Bonn-Cologne Graduate School as well as the from the Deutsche Forschungsgemeinschaft (DFG) via the Sonderforschungsbereich SFB 956. R.E.P is supported by a Discovery grant from NSERC of Canada. R.S.K. acknowledges support from the European Research Council under the European Community’s Seventh Framework Programme (FP7/2007-2013) via the ERC Advanced Grant STARLIGHT (project number 339177) and support from the DFG via the Sonderforschungsbereich SFB 881 {\em The Milky Way System} (subprojects B1, B2, and B5) as well as the Schwerpunktprogramm SPP 1573 {\em Physics of the ISM}. R.S.K also thanks for the warm hospitality at the Department of Astronomy and Astrophysics at the University of California at Santa Cruz and at the Kavli Institute for Particle Astrophysics and Cosmology at Stanford University during a sabbatical stay. The simulations presented here were performed on Supermuc at the Leibniz Supercomputing Centre in Garching and on JUROPA at the Supercomputing Centre in J\"ulich. The FLASH code was developed partly by the DOE-supported Alliances Center for Astrophysical Thermonuclear Flashes (ASC) at the University of Chicago.


\label{lastpage}

\end{document}